%% file: space.tex
\documentclass[a4paper,12pt,oneside]{book}

\usepackage[T1]{fontenc}
\usepackage[ansinew]{inputenc}
\usepackage{amsmath}
\usepackage{amssymb}
\usepackage{graphicx}
\usepackage{color}
\usepackage{euscript}
\usepackage[noautoscale]{youngtab}
\newlength{\defbaselineskip}
\newcommand{\setlinespacing}[1]%
           {\setlength{\baselineskip}{#1 \defbaselineskip}}

\setlinespacing{1.66}
\begin{document}
\include{title}

\frontmatter \normalsize
\setlinespacing{1.66}

 \tableofcontents

\mainmatter

{\typeout{Introduction}
\include{part1-chapter1}}

{\typeout{Null Dynamics in Schwarzschild Spacetime }
\include{part1-chapter2}}

{\typeout{Quantum Gravitational Optics}
\include{part1-chapter3}}

{\typeout{Quantum Modified Trajectories}
\include{part1-chapter4}}

{\typeout{Quantum Modified Schwarzschild Metric}
\include{part1-chapter5}}

{\typeout{Summary}
\include{part1-chapter6}}

{\typeout{Components of the Riemann Tensor in Schwarzschild
Spacetime}
 \include{app1}}

\backmatter

 \setlinespacing{1.50}

\end{document}

%% file: title.tex
\thispagestyle{empty}
 \setlinespacing{1.66}
 \begin{center}
\vskip 2cm \Huge Quantum Modified
Null Trajectories in Schwarzschild Spacetime\\
\vskip 1cm
\Large \vskip 1cm Avtar Singh Sehra
\\

\large
\textit{University of Wales Swansea\\
Singleton Park\\
SA2 8PP\\}
\Large
\vskip 2cm
Abstract
\\
\end{center}
\small
The photon vacuum polarization effect in curved spacetime leads to birefringence, i.e. the photon velocity becomes $c\pm \delta c$ depending on its polarization. This velocity shift then results in modified photon trajectories. We find that photon trajectories are shifted by equal and opposite amounts for the two photon polarizations, as expected by the sum rule \cite{graham2}.  Therefore, the critical circular orbit at $u=1/3M$ in Schwarzschild spacetime, is split depending on polarization as $u=1/3M\pm A\delta(M)$ (to first order in $A$), where $A$ is a constant found to be $\sim 10^{-32}$ for a solar mass blackhole. Then using general quantum modified trajectory equations we find that photons projected into the blackhole for a critical impact parameter tend to the critical orbit associated with that polarization.  We then use an impact parameter that is lower than the critical one.  In this case the photons tend to the event horizon in coordinate time, and according to the affine parameter the photons fall into the singularity.  This means even with the quantum corrections the event horizon behaves in the classic way, as expected from the horizon theorem \cite{graham2}.  We also construct a quantum modified Schwarzschild metric, which encompasses the quantum polarization corrections.  This is then used to derive the photons general quantum modified equations of motion, as before.  When this modified metric is used with wave vectors for radially projected photons we obtain the classic equations of motion, as expected, as radial velocities are not modified by the quantum polarization correction.

%% file: part1-chapter1.tex
\chapter{Introduction}
\label{introduction}
\section{General Relativity}
Since the birth of special relativity, in 1905, the nature of space
and time has been demoted to a relative entity, known as spacetime,
which is stretched and contracted depending on an observer's frame
of reference, while the speed of light, $c$, has taken the pedestal
of an absolute and universal speed limit, unaffected by any
transformation of reference frame. From this emerged a generalized
theory of relativity, which portrayed the gravitational field in a
new and revolutionary way: where it didn't depend on a propagating
field but on the nature of spacetime itself. In this view matter (or
energy) is said to curve and modify the surrounding spacetime, this
then results in photons and particles tracing out shortest paths
between two points, known as geodesics. Therefore, gravitational
forces become a manifestation of the curved spacetime due to the
presence of matter \cite{kenyon,weinberg}. In this general
relativistic framework spacetime is described by the metric
$g_{\mu\nu}$ and the motion of particles are described by the
interval equation:
\begin{equation}
k^2=g_{\mu\nu}k^{\mu}k^{\nu} \qquad \Rightarrow \qquad
                              \begin{array}{c}
                                >0 \qquad \textmd{Time-like ($c<1$)} \\
                                =0 \qquad \textmd{Light-like ($c=0$)} \\
                                <0 \qquad \textmd{Space-like ($c>1$)} \\
                              \end{array}
\end{equation}
where $k^{\mu}=\frac{dx^{\mu}}{d\tau}$\footnote{For photons this
becomes $k^{\mu}=\frac{dx^{\mu}}{d\lambda}$, written in terms of the
affine parameter $\lambda$}.  For flat spacetime or a local inertial
frame (LIF), where $g_{\mu\nu}$ is replaced by the diagonal
Minkowski metric $\eta_{\mu\nu}=(1,-1,-1,-1)$, the interval equation
becomes:
\begin{equation}
k^2=\eta_{\mu\nu}k^{\mu} k^{\nu}=\frac{dt}{d\tau}^2-\frac{d
x}{d\tau}^2 -\frac{d y}{d\tau}^2-\frac{d z}{d\tau}^2
\end{equation}
for $c=1$.  Apart from resolving the problems associated with
Newtonian mechanics, such as describing the perihelion advance of
Mercury, the strongest aspect of general relativity was its
predictive power. One of its most radical claims, and the building
blocks of the theory itself, was that gravitational fields affect
radiation, which was then confirmed through the observation of
starlight deflection by the sun.  From this emerged some profound
and fantastic possibilities such as black holes and gravitational
lensing. The gravitational effect on light rays also leads to the
possibility that photons could follow stable orbits around stars
(discussed in chapter~\ref{null dynamics}), and it's this
possibility in which we will be interested.
\section{QED in a Curved Spacetime}
\label{qed curved space} 
\begin{figure}
    \begin{center}
         \includegraphics[width=0.6\textwidth]{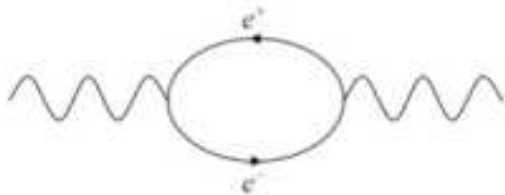}\\
        \caption{First order, $\alpha$, vacuum polarization Feynman diagram contributing to the photon propagator}
        \label{vacuum}
    \end{center}
\end{figure}

Even though photon trajectories are
modified in a curved spacetime, and the resulting curved paths are
described by general relativity, this bending of light was, for a
long time, considered to have no effect on the velocity of the
photon. This view shifted slightly when, in 1980, Drummond and
Hathrell\cite{drummond} proposed that a photon propagating in a
curved spacetime may, depending on its direction and polarization,
travel with a velocity that exceeds the normal speed of light $c$.
This change in velocity would then result in trajectories other then
the ones described by "classical" general relativity. This effect is
simply described as a modification of the light cone in a LIF:
    \begin{equation}
        k^2=\eta_{\mu\nu}k^{\mu}k^{\nu}=0\quad\rightarrow\quad(\eta_{\mu\nu}+\alpha\sigma_{\mu\nu}(R))k^{\mu}k^{\nu}
    \end{equation}
Where $\alpha$ is the fine structure constant and
$\sigma_{\mu\nu}(R)$ is a modification to the metric that depends on
the Riemann curvature at the origin of the LIF.  This correction is
seen to arise from photon vacuum polarizations in a curved space
time, Fig.~\ref{vacuum}.
Qualitatively it can be thought of as a photon splitting into a
virtual $e^+e^-$ pair, so at the quantum level it is characterized
by the Compton wavelength $\lambda_c$; then, when this quantum cloud
of size $\mathcal{O}(\lambda_c)$ passes through a curved spacetime
its motion would be affected differently to that described by
general relativity, possibly in a polarization-dependent way
\cite{graham2}.

This effect of vacuum polarization is considered through the
effective action:
\begin{equation}
S=S_1+S_2 \label{total action}
 \end{equation}
where $S_1$ is the Maxwell electromagnetic action in curved space
time:
\begin{eqnarray}
S_1&=&-\frac{1}{4}\int d^4 x
\sqrt{-g}F_{\mu\nu}F^{\mu\nu}\nonumber\\
F_{\mu\nu}&=&\partial_{\mu}A_{\nu}-\partial_{\nu}A_{\mu}
\label{classic curved action}
\end{eqnarray}
and $S_2$ is the part of the action that incorporates the effects of
virtual electron loops in a background gravitational field.  As we
are only concerned with the propagation of individual photons it
must be quadratic in $A_{\mu}$, and the constraint of gauge
invariance then implies that it must depend on $F_{\mu\nu}$ rather
than $A_{\mu}$.  Also, as the virtual loops give the photon a size
of $\lambda_c$, $S_2$ can be expanded in powers of
$\lambda_c^2=m^{-2}$, thus the lowest term in the expansion would be
of order $m^{-2}$, which is the term corresponding to one electron
loop.  With these constraints there are only four independent gauge
invariant terms, which can be chosen to be:
\begin{eqnarray}
S_2&=&\frac{1}{m^2}\int d^4 x
\sqrt{-g}(aRF_{\mu\nu}F^{\mu\nu}+bR_{\mu\nu}F^{\mu\sigma}F^{\nu}_{\sigma}\nonumber\\
&+&cR_{\mu\nu\sigma\tau}F^{\mu\nu}F^{\sigma\tau}+d
D_{\mu}F^{\mu\nu}D_{\sigma}F^{\sigma}_{\nu}) \label{quantum curved
action}
\end{eqnarray}
The first three terms represent a direct coupling of the
electromagnetic field to the curvature, and they vanish in
flat-spacetime.  The fourth, however, is also applicable in the case
of flat-spacetime, and represents off-mass-shell effects in the
vacuum polarization. In \cite{drummond} the values for $a$, $b$,
$c$, and $d$ have been determined to $\mathcal{O}(e^2)$.  The
constant $d$ is obtained by comparing the coefficient of the
renormalized flat-spacetime photon propagator associated with the
Feynman diagram in Fig.~\ref{vacuum}\footnote{The photon propogator
with the vacuum polarization in flat-spacetime is given (in the
Feynman gauge) by:
$\frac{\eta_{\mu\nu}}{q^2}\rightarrow\frac{\eta_{\mu\nu}}{q^2}+\frac{1}{q^4}I_{\mu\nu}$,
where
$I_{\mu\nu}=(\eta_{\mu\nu}q^2-q_{\mu}q_{\nu})(1-\frac{e^2}{60\pi^2}\frac{q^2}{m_e^2}+\ldots)$
} to the result of the same order given by the effective action $S$;
and $a$, $b$, and $c$ are obtained by comparing the coefficients of
the coupling of a graviton to two photons\footnote{Deduced from the
matrix element $<\gamma(q_2,\beta)T^{\mu\nu}\gamma(q_1,\alpha)>$,
where $T^{\mu\nu}$ is the energy momentum tensor, \cite{drummond}}
to the same result obtained from $S_2$.  In this way the constants
are given as:
\begin{equation}
a=-\frac{5}{720}\frac{\alpha}{\pi} \qquad
b=\frac{26}{720}\frac{\alpha}{\pi} \qquad
c=-\frac{2}{720}\frac{\alpha}{\pi} \qquad
d=-\frac{24}{720}\frac{\alpha}{\pi}
\end{equation}
Then, as the equations of motion for the electromagnetic field are
given by:
\begin{equation}
\frac{\delta S}{\delta A_{\mu}(x)}=0
\end{equation}
using the modified action (\ref{total action}) we find:
\begin{equation}
D_{\mu}F^{\mu\nu}+\frac{\delta S_2}{\delta A_{\mu}}=0
\end{equation}
From this we can see that $D_{\mu}F^{\mu\nu}$ is of
$\mathcal{O}(e^2)$, therefore, the term with coefficient $d$ in
Eqn.~(\ref{quantum curved action}) will be of $\mathcal{O}(e^4)$,
hence we can omit it from the final equation of motion.  In this way
we find \cite{graham}:
\begin{equation}
D_{\mu}F^{\mu\nu}-\frac{1}{m_e^2}[2bR_{\mu\lambda}D^{\mu}F^{\lambda\nu}
+4cg^{\nu\tau}R_{\mu\tau\lambda\rho}D^{\mu}F^{\lambda\rho}]=0
\label{quantum maxwell}
\end{equation}
which is the Maxwell equation in curved spacetime, incorporating the
coupling of curvature with vacuum polarization effects. Using this
modified Maxwell equation and the methods of geometric optics
(described in Chapter~\ref{optics}) it is possible to derive the
quantum modified light cone and geodesic equations.  Then, using
these, we are able to determine the quantum modified trajectories in
curved spacetime.
\section{The Equivalence Principle and Causality}
The equivalence principle exists in two forms: weak and strong. The
weak equivalence principle states that at each point in spacetime
there exists a local Minkowski frame, which is a fundamental
requirement of general relativity\footnote{This implies that general
relativity is formulated on a Riemannian manifold}. The strong
equivalence principle (SEP), on the other hand, states that the laws
of physics are the same in all LIFs at different points in
spacetime, and at the origin of each LIF they take the special
relativistic form. Then the coupling of curvature to the
electromagnetic field in the effective action, Eqn.~(\ref{total
action}), is a violation of the SEP.  Due too this violation of the
SEP, QED in curved spacetime remains a causal theory - despite the
modification to the physical light cone.

\begin{figure}
    \begin{center}
         \includegraphics[width=0.5\textwidth]{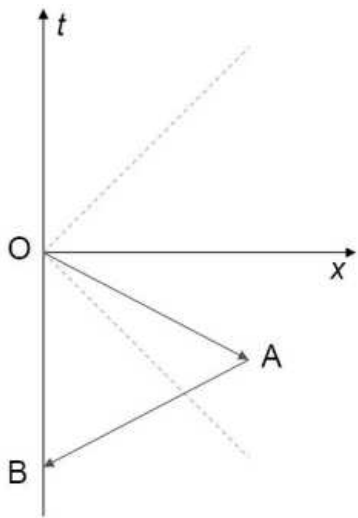}
        \caption{An unacceptable closed-time-loop.}
        \label{time-loop}
    \end{center}
\end{figure}

This can be seen more clearly by considering Global Lorentz
invariance\footnote{Global Lorentz invariance states that the laws
of physics are the same in all inertial frames.} (GLI), which is the
special relativistic equivalent of the SEP.  In special relativity
GLI states that faster than light signals automatically imply the
possibility of unacceptable closed-time-loops, Fig.~\ref{time-loop}.
That is, if you can send a signal backwards in time in one frame,
then it should be possible in any frame.  However, if you break this
GLI, a signal backwards in time in one frame does not automatically
imply you can send a signal backwards in time from any frame.
Therefore, in the case of QED in curved spacetime, due to the
breakdown of the SEP we retain the fundamental property of
causality: faster than light signals can be seen to go backwards in
time in a certain frame, but this no longer implies that you can
send a signal backwards in time from any frame. Thus, in this case,
spacelike motion does not necessarily imply a causality
violation\footnote{Further discussion of causality is given in
\cite{graham3}}.

\section{Critical Stable Orbits} One of the simplest curved
spacetimes is the (Ricci-flat) Schwarzschild spacetime, which
describes the spherically symmetric geometry outside a star. The
geometry of such a spacetime structure is given by the line
interval:
\begin{equation}
  ds^{2}=(1-\frac{2M}{r})dt^{2}-(1-\frac{2M}{r})^{-1}dr^{2}-r^{2}d\theta^{2}-(r^{2}\sin^{2}\theta)d\phi^{2}
  \label{schwarzschild line interval}
\end{equation}
In this spacetime the equations of motion, derived for $k^2=0$, have
a special solution for null geodesics.  This solution describes a
light ray in a circular orbit with a radius $r=3M$ and an associated
critical impact parameter.  Therefore, a photon projected from
$r\rightarrow\infty$ with the critical parameter tends to the stable
circular orbit by spiralling around it.  But, in the context of
vacuum polarization effects, we may derive the null equation of
motion using the quantum modified Maxwell equation.  In this way the
orbit equation should give us stable circular orbits that are
dependent on the photon polarization, i.e. splitting the circular
orbit depending on polarization. Therefore, a photon coming in from
$r\rightarrow\infty$, with a particular polarization and impact
parameter, should tend to its associated critical stable orbit by
spiralling around it in the classic way.

%% file: part1-chapter2.tex
\chapter{Null Dynamics in Schwarzschild Spacetime }
\label{null dynamics}  In this chapter we will give a brief overview
of null dynamics in Schwarzschild spacetime.  Starting with the
geodesic equations and the line interval for photons we will derive
the orbit equation, which will then be used to determine the
critical stable orbit for photons and the associated impact
parameter.  We will then go on to show that decreasing the impact
parameter, from the critical value, causes the photon trajectory,
given as a function of $\phi$, to spiral into the singularity, but
as a function of $t$ it tends to the event horizon.  The main aim of
this chapter is to familiarise ourselves with the classical
solutions of photon trajectories in Schwarzschild spacetime, so we
can then compare the equivalent results for the quantum modified
case.
\section{Equations of Motion}
When we consider motion in a plane, where $\theta=\frac{\pi}{2}$ and
$\dot{\theta}=\ddot{\theta}=0$, the geodesic equations for
Schwarzschild spacetime, (\ref{equation of motion t})-(\ref{equation
of motion
 phi}), can written as:
    \begin{equation}
        (1-\frac{2M}{r})\ddot{t}+\frac{2M}{r^2}\dot{r}\dot{t}=0
         \label{t with theta=0}
    \end{equation}
    \begin{equation}
        \frac{M}{r^2}\dot{t}^2-r \dot{\phi}^{2} +(1-\frac{2M}{r})^{-1}\ddot{r}
        -\frac{M}{r^{2}}(1-\frac{2M}{r})^{-2}\dot{r}^{2}=0
         \label{r with theta=0}
    \end{equation}
    \begin{equation}
        2r\dot{r} \dot{\phi}+r^2 \ddot{\phi}=0
         \label{phi with theta=0}
    \end{equation}
And the spacetime line interval becomes:
    \begin{equation}
        (1-\frac{2M}{r})\dot{t}^{2}-(1-\frac{2M}{r})^{-1}\dot{r}^{2}-r^{2}\dot{\phi}^{2}=K
         \label{interval with theta=0}
    \end{equation}
where $K=0$ for photons and $\pm1$ for space-like or time-like
motion. Now, to derive the orbit equation $\frac{du}{d\phi}$ we use
Eqns.~(\ref{t with theta=0}), (\ref{phi with theta=0}) and the
interval equation (\ref{interval with theta=0}), in the simplified
form:
 \begin{eqnarray}
\label{equation of motion t theta=0}
        0&=&\ddot{t}+\frac{2M}{r^2}(1-\frac{2M}{r})^{-1}\dot{r}\dot{t}\\
         \label{equation of motion phi theta=0}
        0&=&\ddot{\phi}+\frac{2\dot{r}}{r} \dot{\phi}\\
 \label{interval equation theta=0}
        K(1-\frac{2M}{r})&=&-\dot{r}^{2}+(1-\frac{2M}{r})^{2}\dot{t}^{2}-r^{2}(1-\frac{2M}{r})\dot{\phi}^{2}
         \end{eqnarray}
In order to specify photon trajectories we can set $K=0$ at anytime,
however, then $\tau$ would be interpreted as an affine parameter and
not proper time. To proceed we divide~(\ref{equation of motion t
theta=0}) by $\frac{dt}{d\tau}$ and~(\ref{equation of motion phi
theta=0}) by $\frac{d\phi}{d\tau}$:
    \begin{eqnarray}
        0&=&\frac{\ddot{t}}{\dot{t}}+\frac{2M}{r^2}(1-\frac{2M}{r})^{-1}\dot{r}\\
        0&=&\frac{ \ddot{\phi}}{\dot{\phi}}+\frac{2\dot{r}}{r}
    \end{eqnarray}
These can written as:
    \begin{eqnarray}
        0&=&\frac{d}{d\tau}(\ln{\dot{t}}+\ln(1-\frac{2M}{r}))\\
        0&=&\frac{d}{d\tau}(\ln{\dot{\phi}}+\frac{2\dot{r}}{r})
    \end{eqnarray}
Then, solving these we have:
    \begin{eqnarray}
 \label{dot-t}
        \dot{t}&=&(1-\frac{2M}{r})^{-1}E\\
   \label{dot-phi}
        \dot{\phi}&=&\frac{J}{r^2}
    \end{eqnarray}
Where $E$ and $J$ are constants of integration denoting total energy
and angular momentum about an axis normal to the plane
$\theta=\pi/2$. Now, using Eqns.~(\ref{dot-t}) and (\ref{dot-phi})
to substitute for $\dot{t}$ and $\dot{\phi}$ in Eqn~(\ref{interval
equation theta=0}) and rearranging, we have:
    \begin{equation}
        \dot{r}^2+r^2(1-\frac{2M}{r})\frac{J^2}{r^4}+K(1-\frac{2M}{r})=(1-\frac{2M}{r})^{2}(1-\frac{2M}{r})^{-1}E^2
         \label{interval with dot-t and dot-phi replaced}
    \end{equation}
Rearranging and simplifying, this becomes:
    \begin{eqnarray}
  \label{dr/dtau-1}
        (\frac{dr}{d\tau})^2+(1-\frac{2M}{r})(K+\frac{J^2}{r^2})=E^2\\
        (\frac{dr}{d\tau})=E[1-(1-\frac{2M}{r})(\frac{K}{E^2}+\frac{J^2}{E^{2}r^2})]^{\frac{1}{2}}
    \end{eqnarray}
We now have the components of the four momentum $P^{\alpha}$ in
spherical polar coordinates
    \begin{eqnarray}
        P^{\alpha}&=&(\dot{t},\dot{r},\dot{\theta},\dot{\phi})\nonumber\\
        &=&E(F^{-1},[1-F(\frac{K}{E^2}+\frac{D^2}{r^2})]^{\frac{1}{2}},0,\frac{D}{r^2})
         \label{four momentum}
    \end{eqnarray}
where $D=\frac{J}{E}$ is the impact parameter, $F=(1-\frac{2M}{r})$
and $K=0$ for photons and $K=+1$ for particles.
\subsection{Orbit Equation}
In order to derive orbit equations that are physically
understandable we need to represent them as $r(\phi)$, $r(t)$ and
$\phi(t)$. In this form we can analyse the orbit paths as a function
of rotational angle $\phi$ and the time taken to reach a certain
point along the angle. Therefore, by using
    \begin{equation}
        \frac{dr}{d\tau}=\frac{dr}{d\phi}\cdot\frac{d\phi}{d\tau}=\frac{dr}{d\phi}\frac{J}{r^2}
         \label{dr/dtau=diff}
    \end{equation}
    we can rewrite~(\ref{dr/dtau-1}) as
    \begin{eqnarray}
        E^2&=&(\frac{dr}{d\phi})^2\frac{J^2}{r^4}+(1-\frac{2M}{r})(K+\frac{J^2}{r^2})\nonumber\\
     \Rightarrow \qquad  (\frac{dr}{d\phi})^2&=&(E^2-K)\frac{r^4}{J^2}+2MK\frac{r^3}{J^2}-r^2+2Mr
         \label{dr/dphi}
    \end{eqnarray}
Now, transforming $u=r^{-1}$, so at $r=\infty$ we have $u=0$,
Eqn.~(\ref{dr/dphi}) becomes:
    \begin{eqnarray}
 \label{du/dr with u=1/r}
        (\frac{du}{d\phi})^2r^4&=&(E^2-K)\frac{r^4}{J^2}+2MK\frac{r^3}{J^2}-r^2+2Mr\nonumber\\
  \label{du/dphi}
        (\frac{du}{d\phi})^2&=&\frac{(E^2-K)}{J^2}+2MK\frac{u}{J^2}-u^2+2Mu^3
   \end{eqnarray}
Doing similar manipulation for $\phi(t)$ we have
    \begin{equation}
        \frac{d\phi}{dt}=\frac{d\phi}{d\tau}\cdot\frac{d\tau}{dt}
         \label{dphi/dt=diff}
    \end{equation}
    and using Eqns.~(\ref{dot-t}) and (\ref{dot-phi}) in this, and transforming $u=\frac{1}{r}$, we
    have:
    \begin{equation}
        \frac{d\phi}{dt}=\frac{D}{r^2}(1-\frac{2M}{r})=Du^2(1-2Mu)
         \label{dphi/dt}
    \end{equation}
Finally, we can determine $u(t)$ by inserting:
    \begin{equation}
        \frac{du}{d\phi}=\frac{du}{dt}\cdot\frac{dt}{d\phi}=\frac{du}{dt}[D u^2(1-2M
        u)]^{-1}
         \label{du/dphi=diff}
    \end{equation}
into Eqn.~(\ref{du/dphi}), which gives
    \begin{equation}
        (\frac{du}{dt})^2=[Du^2(1-2Mu)]^2[\frac{(E^2-K)}{J^2}+2MK\frac{u}{J^2}-u^2+2Mu^3]
         \label{du/dt}
    \end{equation}
Now, the Eqns.~(\ref{du/dphi}), (\ref{dphi/dt}) and (\ref{du/dt})
are the general equations that determine photon and particle
trajectories in the plane $\theta=\pi/2$. If we take $K=0$ we then
have the required photon trajectory equations
    \begin{eqnarray}
\label{du/dphi=f(u)}
        (\frac{du}{d\phi})^2&=&\frac{1}{D^2}-u^2+2Mu^3=f(u)\\
 \label{du/dt=f(u)}
        (\frac{du}{dt})^2&=&[Du^2(1-2Mu)]^2[\frac{1}{D^2}-u^2+2Mu^3]\nonumber\\
        &=&[Du^2(1-2Mu)]^2f(u)
    \end{eqnarray}
We can also write another equation, specifically for null radial
geodesics.  By using the fact that $\dot{\phi}=0$,
Eqn.~(\ref{dot-phi}) implies $J=0$, then Eqns~(\ref{dot-t}) and
(\ref{dr/dtau-1}) become
    \begin{eqnarray}
        (1-\frac{2M}{r})\frac{dt}{d\tau}=E\\
        \frac{dr}{d\tau}=\pm E
         \label{dr/dtau J=0}
    \end{eqnarray}
    Combining these we have:
    \begin{equation}
        \frac{dr}{dt}=\pm (1-\frac{2M}{r})
         \label{dr/dt J=0}
    \end{equation}
\section{Orbits in Schwarzschild Spacetime}
We will now discuss the solutions of Eqns.~(\ref{du/dphi=f(u)}),
(\ref{du/dt=f(u)}) and (\ref{dr/dt J=0}).   We will start off with
the simple case of the radial geodesic and then go onto the case of
the general orbits. For the general case we will solve
(\ref{du/dphi=f(u)}) and (\ref{du/dt=f(u)}) to determine the
critical stable orbits and the associated impact parameters.  We
will then go on to study the trajectories of photons as they fall
into the singularity
\subsection{Null Radial Geodesic}
\begin{figure}
    \begin{center}
        \includegraphics[width=0.8\textwidth]{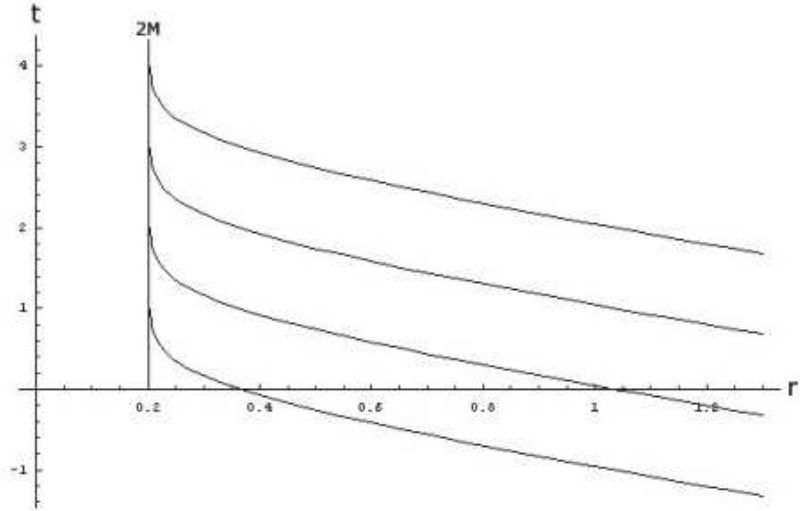}\\
        \caption{Radial Null geodesics take an infinite coordinate
         time to reach the event horizon at 2M}\label{coordinate-radial}
    \end{center}
\end{figure}

\begin{figure}
    \begin{center}
        \includegraphics[width=0.8\textwidth]{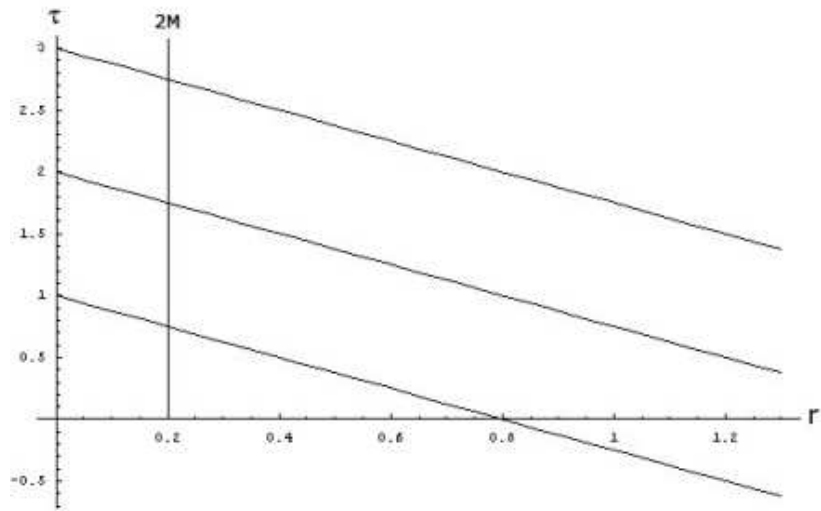}\\
        \caption{Radial Null geodesics reach and cross the event horizon
        at 2M without noticing it when defined with the affine parameter}\label{proper-radial}
    \end{center}
\end{figure}

Eqn.~(\ref{dr/dt J=0}) can be solved by rewriting it as:
    \begin{eqnarray}
  \label{rearranged dr/dt J=0}
        \frac{dt}{dr}&=&\pm (1-\frac{2M}{r})^{-1}\\
    \label{coordinate time solution}
        \Rightarrow \qquad t&=&\pm(r+2M\log(\frac{r}{2M}-1))+ constant_{\pm}
    \end{eqnarray}
This solution (specifically the $-$ one) gives the trajectory of a
photon coming from infinity and into the black hole.  It shows that
the photon takes an infinite coordinate time to reach the horizon,
which can be seen in Fig.~\ref{coordinate-radial}. However, if we
solve Eqn.~(\ref{dr/dtau J=0}), i.e. in terms of the affine
parameter, we can show that the photon reaches and crosses the event
horizon without ever noticing it,
    \begin{equation}
        r=\pm E \tau + constant_{\pm}
        \label{proper time solution}
    \end{equation}
this can be seen in Fig.~\ref{proper-radial}
\subsection{General Null Geodesics and Critical Orbits}
We will first solve Eqn.~(\ref{du/dphi=f(u)}) in order to determine
the the critical stable photon orbits.  In order to do this we first
consider the point of equilibrium and the associated impact
parameter.  This equilibrium point occurs when
    \begin{equation}
        \frac{du}{d\phi}=0
        \label{du/dphi=0}
    \end{equation}
which means as the photon is orbiting the black hole $u$ does not
change; so if the radial distance does not change it implies a
circular orbit. Therefore we must solve
    \begin{equation}
         f(u)=\frac{1}{D^2}-u^2+2Mu^3=0
         \label{f(u)=0}
    \end{equation}
The sum and product of the roots $u_1$, $u_2$, and $u_3$ of this
equation are given by\footnote{for $au^3+bu^2+c=0$ we have
$u_1+u_2+u_3=-\frac{b}{a}$ and $u_1u_2u_3=-\frac{c}{a}$}
    \begin{equation}
        u_1+u_2+u_3=\frac{1}{2M} \qquad \textrm{and} \qquad u_1u_2u_3=-\frac{1}{2MD^2}
        \label{root-condition}
    \end{equation}
This shows that $f(u)=0$ must have a real negative root, and the two
remaining roots can be real (distinct or coincident) or be a complex
conjugate pair; however, the occurrence of coincident positive real
roots implies the existence of a circular orbit. Thus, if a
coincident root occurs it should be at the point given by the
derivative of $f(u)$
    \begin{equation}
         f'(u)=6Mu^2-2u=0
         \label{f'(u)=0}
    \end{equation}
Which then has the solution $u_1=u_2=(3M)^{-1}$.  For this solution
the impact parameter of equation~(\ref{f(u)=0}) is $D=(3\sqrt{3})M$.
From the product condition, equation~(\ref{root-condition}), we find
that the roots of $f(u)=0$ are
    \begin{equation}
        u_1=u_2=\frac{1}{3M} \qquad \textrm{and} \qquad
        u_3=-\frac{1}{6M} \qquad \textrm{and} \qquad D=(3\sqrt{3})M
        \label{classical roots of orbits}
    \end{equation}
Therefore, when the impact parameter is $D=(3\sqrt{3})M$ then
$\frac{du}{d\phi}$ vanishes for $u=(3M)^{-1}$, which implies a
circular orbit of radius $3M$ is an allowed null geodesic
\cite{chandrasekhar}.

\begin{figure}
    \begin{center}
        \includegraphics[width=0.9\textwidth]{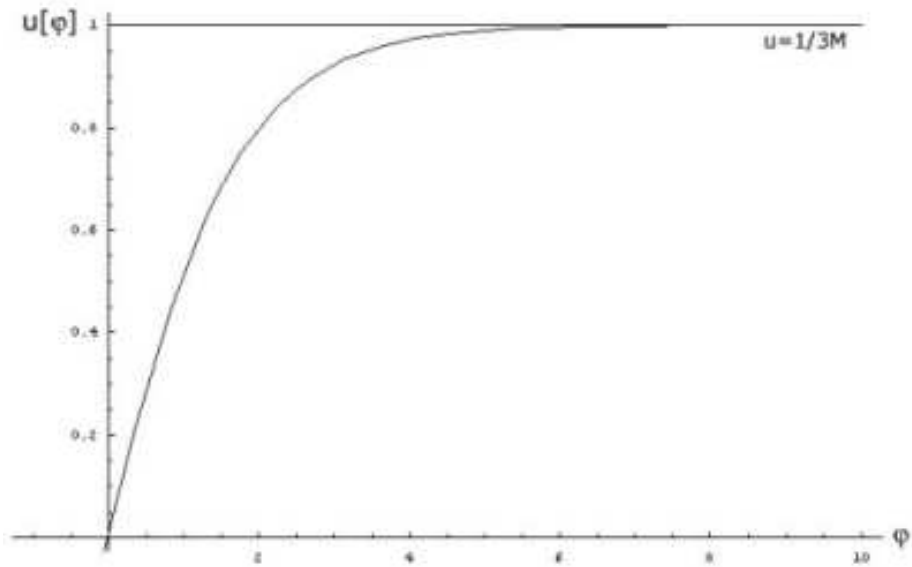}\\
        \caption{Null geodesic, with impact parameter $D=(3\sqrt{3})M$,
        arriving from infinity and approaching $u=\frac{1}{3M}$ asymptotically (M=1/3)}
        \label{u vs phi with D=critical}
    \end{center}
\end{figure}

\begin{figure}
    \begin{center}
        \includegraphics[width=0.9\textwidth]{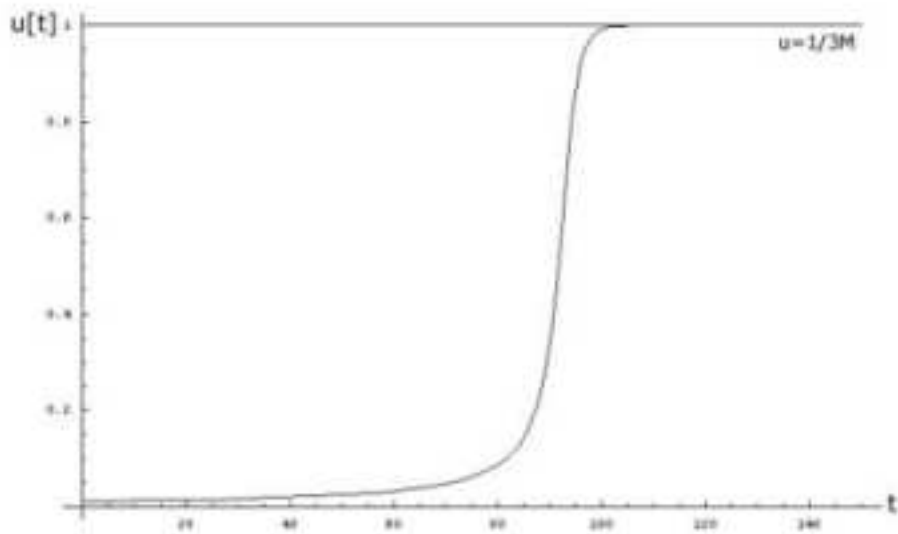}\\
        \caption{For impact parameter $D=(3\sqrt{3})M$ a null trajectory tends to
        $u=\frac{1}{3M}$ asymptotically with time.}
        \label{u vs t with D=critical}
    \end{center}
\end{figure}

Now we can consider a photon at $u=0$ with an impact parameter
$D=(3\sqrt{3})M$.  This, then, gives a trajectory of a photon
spiralling in and tending to the critical orbit at $u=(3M)^{-1}$.
The general differential equation for this impact parameter is given
by rearranging and substituting for $D$ in Eqn.~(\ref{du/dphi=f(u)})
    \begin{equation}
        (\frac{du}{d\phi})^2=2M(u+\frac{1}{6M})(u-\frac{1}{3M})^2
        \label{du/dphi for D=critical}
    \end{equation}
From \cite{chandrasekhar} we have the solution to this as
    \begin{equation}
        u=-\frac{1}{6M}+\frac{1}{2M}\tanh^2{\frac{1}{2}(\phi-\phi_0)}
        \label{u=critical sol}
    \end{equation}
where $\phi_0$ is a constant of integration, given by:
    \begin{equation}
        \tanh^2{(-\frac{1}{2}\phi_0)}=\frac{1}{3},
    \end{equation}
which gives: $u=0$ ($r\rightarrow\infty$) when $\phi=0$, and
$u=\frac{1}{3M}$ when $\phi\rightarrow\infty$. Therefore a null
geodesic arriving from infinity with an impact parameter
$D=(3\sqrt{3})M$ approaches the circle of radius $3M$,
asymptotically, by spiralling around it, as can be seen in
Fig.~\ref{u vs phi with D=critical}\footnote{This figure was plotted
using Eqn~(\ref{u=critical sol}).  We also obtained the same plot by
numerically solving Eqn~(\ref{du/dphi for D=critical}) in
Mathematica.}. Also, numerically solving Eqn~(\ref{du/dt=f(u)}) we
can show that as time increases $u$ tends to $\frac{1}{3M}$, which
can be seen in Fig.~\ref{u vs t with D=critical}.

Finally we can show that when an impact parameter other than
$D=(3\sqrt{3})M$ is used, for example if we set
$D=(3\sqrt{3})M-0.1$, the solution of Eqn.~(\ref{du/dphi=f(u)})
shows that the photon falls past the critical orbit, through the
event horizon, and into the singularity, Fig.~\ref{u vs phi with
D=critical-0.1}. Using this new impact parameter in
Eqn.~(\ref{du/dt=f(u)}) and, again solving numerically, we see that
the photon comes in from infinity and tends to the event horizon,
$u=\frac{1}{2M}$, asymptotically in time $t$, Fig.~\ref{u vs t with
D=critical-0.1}.
\begin{figure}
    \begin{center}
        \includegraphics[width=0.9\textwidth]{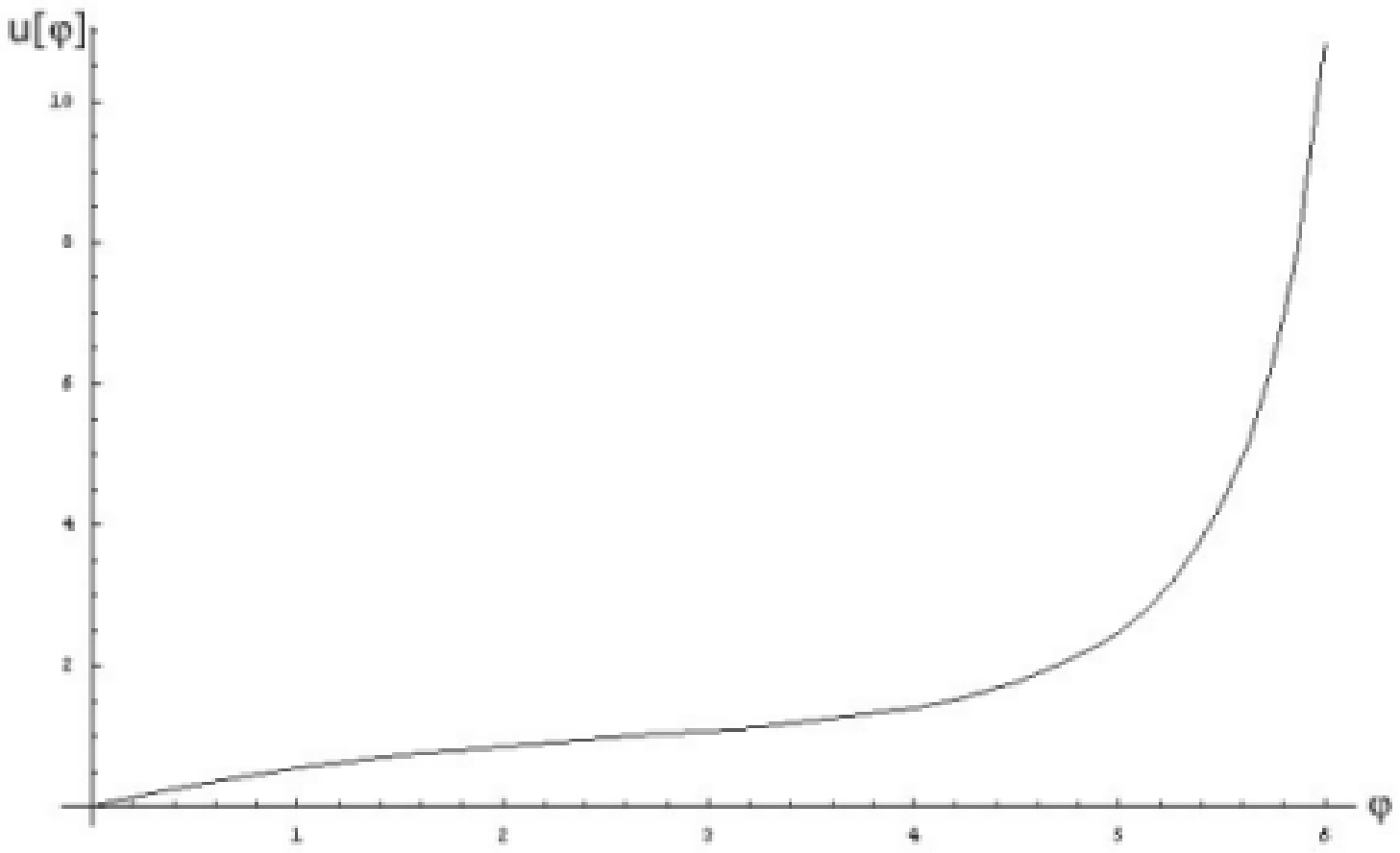}\\
        \caption{Null geodesic, with impact perimeter
        $D=(3\sqrt{3})M-0.1$, arrives from infinity and spirals into the singularity.}
        \label{u vs phi with D=critical-0.1}
    \end{center}
\end{figure}
\begin{figure}
    \begin{center}
        \includegraphics[width=0.9\textwidth]{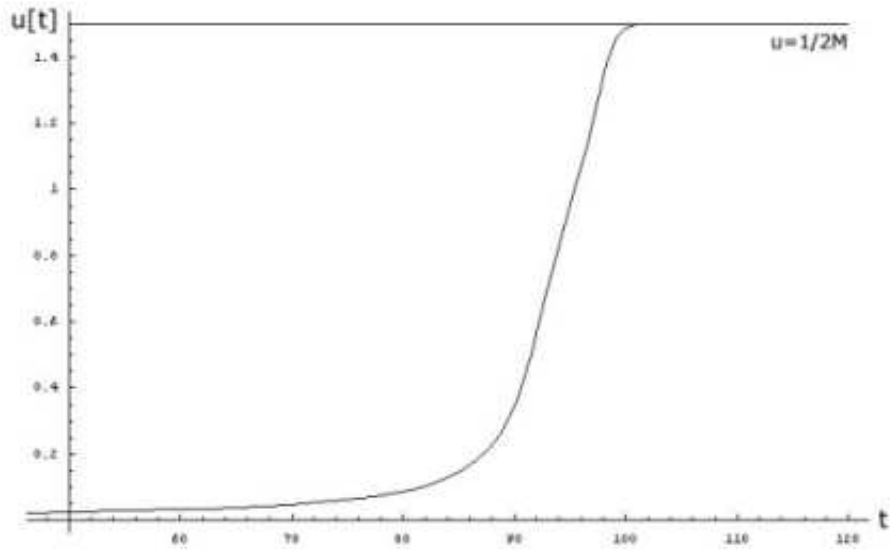}\\
        \caption{Null geodesic, with impact perimeter
        $D=(3\sqrt{3})M-0.1$, arrives from infinity and tends to $u=\frac{1}{2M}$ asymptotically in time $t$.}
        \label{u vs t with D=critical-0.1}
    \end{center}
\end{figure}

%% file: part1-chapter3.tex
\chapter{Quantum Gravitational Optics}
\label{optics}
\section{Photon Propagation in Curved Spacetime}
Maxwell's equations, in curved spacetime,
\begin{equation}
0=D_{\mu}F^{\mu\nu} \label{max1}
\end{equation}
\begin{eqnarray}
\label{max2}
0&=&D_{\mu}F_{\nu\lambda}+D_{\nu}F_{\lambda\mu}+D_{\lambda}F_{\mu\nu}\\
\label{elect-vec} &\Rightarrow&
F_{\mu\nu}=\partial_{\mu}A_{\nu}-\partial_{\nu}A_{\mu},
\end{eqnarray}
cannot be solved explicitly, and even in cases of extreme symmetry
explicit solutions are difficult; this is due to the fact that as
curved space acts as a dispersive material (i.e. bending light rays)
plane wave solutions do not exist.
\subsection{Geometric Optics in Curved Spacetime}
In dispersive materials, where light rays are bent, we can consider
the solution of Maxwell's equations to be a simple perturbation of
the plane wave solution. For example in curved space, relative to an
observer, the electromagnetic waves can appear to be plane and
monochromatic on a scale that is much larger compared to the typical
wavelength, but very small compared with the typical radius of
curvature of space time. Such "locally plane" waves can be
represented, in geometric optics, by approximate solutions of
Maxwell's equations of the form\cite{schneider}:
\begin{equation}
\mathcal{F}^{\mu\nu}=Re(F_{1}^{\mu\nu}+i\varepsilon
F_{2}^{\mu\nu}+\ldots)e^{i\frac{\theta}{\varepsilon}} \label{faraday
tensor}
\end{equation}
and the electromagnetic field vector, defined by
Eqn.~(\ref{elect-vec}), takes the form:
\begin{equation}
\mathcal{A}^{\mu}=Re(A_1^{\mu}+i\varepsilon
A_2^{\mu}+\ldots)e^{i\frac{\theta}{\varepsilon}}
\end{equation}
where the electromagnetic field is written as a slowly-varying
amplitude and a rapidly-varying phase. The parameter $\varepsilon$
is introduced in order to keep track of the relative order of
magnitude of terms, so in curved space Maxwell's equations can be
solved order-by-order in $\varepsilon$.  In this formulation the
wave vector is then defined as the gradient of the phase of the
field,
$k_{\mu}=D_{\mu}\frac{i}{\varepsilon}\theta=\partial_{\mu}\frac{i}{\varepsilon}\theta$,
which in terms of the quantum interpretation is identified as the
photon momentum. We can also write $A^{\mu}=Aa^{\mu}$, where $A$
represents the amplitude and $a_{\mu}$ (normalized as
$a_{\mu}a^{\mu}=-1$) specifies the wave polarization.  These vectors
then satisfy the condition $k_{\mu}a^{\mu}=0$.
\subsubsection{Geometric Optics and Null Dynamics}
In this notation Eqn.~(\ref{max1}) can be written, to leading order
$\mathcal{O}(\frac{1}{\varepsilon})$, as:
\begin{eqnarray}
\partial_{\mu}(F_1^{\mu\nu}e^{i\frac{\theta}{\varepsilon}})&=&
F_1^{\mu\nu}e^{i\frac{\theta}{\varepsilon}}(\frac{i}{\varepsilon}\partial_{\mu}\theta)\nonumber\\
\Rightarrow \qquad k_{\mu}F_1^{\mu\nu}&=&0
\end{eqnarray}
and Eqn.~(\ref{elect-vec}) becomes:
\begin{eqnarray}
F_{1\mu\nu} e^{i\frac{\theta}{\varepsilon}}&=&(\partial_{\mu}A_{1\nu}-\partial_{\nu}A_{1\mu})e^{i\frac{\theta}{\varepsilon}}\nonumber\\
&=&\frac{i}{\varepsilon}(\partial_{\mu}\theta
A_{1\nu}-\partial_{\nu}\theta
A_{1\mu})e^{i\frac{\theta}{\varepsilon}}\nonumber\\
\Rightarrow \qquad F_{1\mu\nu}&=&k_{\mu}A_{1\nu}-k_{\nu}A_{1\mu}
\label{elect-vect-geometric}
\end{eqnarray}
Now, by combining these, we have:
\begin{eqnarray}
k_{\mu}F_1^{\mu\nu}&=&k_{\mu}(k^{\mu}A_1^{\nu}-k^{\nu}A_1^{\mu})=0\nonumber\\
&=&(k_{\mu}k^{\mu})A_1^{\nu}-k^{\nu} (k_{\mu}A_1^{\mu})\nonumber\\
 \Rightarrow \qquad k^2a^{\nu}&=&0
 \end{eqnarray}
and from this we can deduce that $k^2=0$, i.e. $k^{\mu}$ is a null
vector.  Also, it follows from the definition of $k^{\mu}$ as a
gradient that $D_{\mu}k_{\nu}=D_{\nu}k_{\mu}$, so
\begin{equation}
k^{\mu}D_{\mu}k^{\nu}=k^{\mu}D^{\nu}k_{\mu}=\frac{1}{2}D^{\nu}k^2=0
\label{condition for geodesic}
\end{equation}
Using this, and the fact that light rays are defined as the curves
given by $x_{\mu}(s)$ where $\frac{dx_{\mu}}{ds}=k_{\mu}$, we can
derive the geodesic equation as follows\cite{graham}:
\begin{eqnarray}
0&=&k^{\mu}D_{\mu}k^{\nu}\nonumber\\
&=&\frac{d^2x^{\nu}}{ds^2}+\Gamma^{\nu}_{\mu\lambda}\frac{dx^{\mu}}{ds}\frac{dx^{\lambda}}{ds}
\end{eqnarray}
\section{Quantum Modified Null Dynamics}
As was seen in Sec.~\ref{qed curved space}, using the effective
action, Eqn.~(\ref{total action}), the equation of motion becomes:
\begin{equation}
0=D_{\mu}F^{\mu\nu}-\frac{1}{m^2}[2bR_{\mu\lambda}D^{\mu}F^{\lambda\nu}+4cg^{\nu\tau}R_{\mu\tau\lambda\rho}D^{\mu}F^{\lambda\rho}]
\label{quantum modified maxwell equation}
\end{equation}
and the Bianchi identity, Eqn.~(\ref{max2}), remains unchanged.
Now, as before we can determine the quantum modified light cone and
geodesic equations.
\subsubsection{Quantum Modified Light Cone}
Substituting Eqn.~(\ref{faraday tensor}) in (\ref{quantum modified
maxwell equation}) we find, again to
$\mathcal{O}(\frac{1}{\varepsilon})$:
\begin{eqnarray}
0=\frac{i}{\varepsilon}(D_{\mu}\theta)e^{i\frac{\theta}{\varepsilon}}
F_1^{\mu\nu}&-&\frac{1}{m^2}[2bR_{\mu\lambda}\frac{i}{\varepsilon}(D^{\mu}\theta)e^{i\frac{\theta}{\varepsilon}}
F_1^{\lambda\nu}\nonumber\\
&+&4cg^{\nu\tau}R_{\mu\tau\lambda\rho}\frac{i}{\varepsilon}(D^{\mu}\theta)e^{i\frac{\theta}{\varepsilon}}
F_1^{\lambda\rho}]\nonumber
\end{eqnarray}
\begin{eqnarray}
\Rightarrow \qquad
0=k_{\mu}F_1^{\mu\nu}-\frac{1}{m_e^2}[2bR_{\mu\lambda}k^{\mu}F_1^{\lambda\nu}+4cg^{\nu\tau}R_{\mu\tau\lambda\rho}k^{\mu}F_1^{\lambda\rho}]
\end{eqnarray}
and using Eqn.~(\ref{elect-vect-geometric}) we have:
\begin{eqnarray}
 0=k_{\mu}( k^{\mu}A_1^{\nu}-k^{\nu}A_1^{\mu})
&-&\frac{1}{m_e^2}[2bR_{\mu\lambda}k^{\mu}
(k^{\lambda}A_1^{\nu}-k^{\nu}A_1^{\lambda})\nonumber\\
&+&4cg^{\nu\tau}R_{\mu\tau\lambda\rho}k^{\mu}
(k^{\lambda}A_1^{\rho}-k^{\rho}A_1^{\lambda}) ]\nonumber
\end{eqnarray}
using $k_{\mu}A^{\mu}=0$ this becomes:
\begin{eqnarray}
0=k_{\mu}k^{\mu}A_1^{\nu}&-& \frac{1}{m_e^2}[2bR_{\mu\lambda}k^{\mu}
(k^{\lambda}A_1^{\nu}-k^{\nu}A_1^{\lambda})\nonumber\\
&+&4cg^{\nu\tau}R_{\mu\tau\lambda\rho}k^{\mu}
(2k^{\lambda}A_1^{\rho}) ]
\end{eqnarray}
where $A_1^{\nu}=A_1a^{\nu}$, and the last line is simplified by
relabeling of indices. Now, Contracting with $Aa_{\nu}$ and
eliminating $A$ we have:
\begin{eqnarray} 0=-k_{\mu}k^{\mu}+ \frac{2b}{m_e^2}R_{\mu\lambda}k^{\mu}
k^{\lambda}-\frac{8c}{m_e^2}R_{\mu\tau\lambda\rho}k^{\mu}
k^{\lambda}a_1^{\tau}a_1^{\rho}\nonumber
\end{eqnarray}
This then gives the quantum modified light cone:
\begin{eqnarray}
k^2-\frac{2b}{m^2}R_{\mu\lambda}k^{\mu} k^{\lambda}
-\frac{8c_{\alpha}}{m_{e}^2}R_{\mu\tau\lambda\rho}k^{\mu}
k^{\lambda}a^{\tau}a^{\rho}=k^2-\delta k(a)=0 \label{quantum
modified light cone equation}
    \end{eqnarray}
where we have replaced the constant $c=-\frac{\alpha}{360\pi}$ with
$c_{\alpha}=\frac{\alpha}{360\pi}$ for convenience of
interpretation, i.e. the sign of the light cone is immediately
obvious from $R_{\mu\tau\lambda\rho}k^{\mu}
k^{\lambda}a^{\tau}a^{\rho}$. As we will be working in the
Schwarzschild spacetime the quantum modified light cone for the
Ricci flat case ($R_{\mu\lambda}=0$) is given by:
\begin{equation}
k^2=\frac{8c_{\alpha}}{m_{e}^2}R_{\mu\tau\lambda\rho}k^{\mu}
k^{\lambda}a^{\tau}a^{\rho}=\delta k (a) \label{quantum modified
light cone correction}
\end{equation}
Here the sign of the light cone depends on polarization and the
photon trajectory; if the correction is positive we have space-like
motion, and if it's negative we have time-like motion.
\subsubsection{Quantum Modified Geodesic Equation}
The photon trajectories corresponding to the quantum modified
equation of motion, (\ref{quantum modified maxwell equation}), can
be represented by a generalised version of Eqn.~(\ref{condition for
geodesic}):
\begin{eqnarray}
0&=&\frac{1}{2}D_{\nu}[ k^2-\frac{2b}{m^2}R_{\mu\lambda}k^{\mu}
k^{\lambda}
-\frac{8c_{\alpha}}{m_{e}^2}R_{\mu\tau\lambda\rho}k^{\mu}
k^{\lambda}a^{\tau}a^{\rho}]\nonumber\\
0&=&\frac{1}{2}D_{\nu}k^2-\frac{1}{m^2}D_{\nu}[bR_{\mu\lambda}k^{\mu}
k^{\lambda} +4c_{\alpha}R_{\mu\tau\lambda\rho}k^{\mu}
k^{\lambda}a^{\tau}a^{\rho}]\nonumber\\
\Rightarrow \qquad
0&=&\frac{d^2x^{\nu}}{ds^2}+\Gamma_{\mu\lambda}^{\nu}\frac{dx^\mu}{ds}\frac{dx^\lambda}{ds}\nonumber\\
&-&\frac{1}{m^2}\partial_{\nu}[(bR_{\beta\gamma}+4c_{\alpha}R_{\beta\sigma\gamma\tau}a^\sigma
a^\tau)\frac{dx^\beta}{ds}\frac{dx^\gamma}{ds}]
    \label{quantum modified geodesic equation}
\end{eqnarray}
where we have used $k^{\mu}=\frac{dx^{\mu}}{ds}$, and covariant
derivative in the second term is replaced by a partial derivative as
it's acting on a scalar.  In Ricci spacetime this equation becomes:
\begin{equation}
0=\frac{d^2x^{\nu}}{ds^2}+\Gamma_{\mu\lambda}^{\nu}\frac{dx^\mu}{ds}\frac{dx^\lambda}{ds}-\frac{4c_{\alpha}}{m^2}\partial_{\nu}[R_{\beta\sigma\gamma\tau}a^\sigma
a^\tau\frac{dx^\beta}{ds}\frac{dx^\gamma}{ds}]
    \label{quantum modified geodesic equation ricci}
\end{equation}
\section{Horizon Theorem and Polarization Rule}
There are two general features associated with quantum modified
photon propagation\cite{graham}.  First, it is a general result that
the velocity of radially directed photons remains equal to $c$ at
the event horizon. Second, for Ricci flat spacetimes (such as
Schwarzschild\cite{drummond} and Kerr\cite{daniels} spacetimes), the
velocity shifts for the two transverse polarizations are always
equal and opposite.  However, this is no longer true for non-Ricci
flat cases (such as Robertson-Walker spacetime\cite{drummond}).  In
these cases, the polarization averaged velocity shift is
proportional to the matter energy-momentum tensor. These features
can be easily shown by using the Newman-Penrose formalism: this
characterises spacetimes using a set of complex scalars, which are
found by contracting the Weyl tensor with elements of a null
tetrad\cite{chandrasekhar}.
\subsubsection{Newman-Penrose Formalism}
We choose the basis vectors of the null tetrad as\cite{graham}:
$l^{\mu}=k^{\mu}$, the photon momentum.  Then, we denote the two
spacelike, normalized, transverse polarization vectors by $a^{\mu}$
and $b^{\mu}$ and construct the null vectors
$m^{\mu}=\frac{1}{\sqrt{2}}(a^{\mu}+ib^{\mu})$ and
$\bar{m}^{\mu}=\frac{1}{\sqrt{2}}(a^{\mu}-ib^{\mu})$.  We complete
the tetrad with a further null vector $n^{\mu}$, which is orthogonal
to $m^{\mu}$ and $\bar{m}^{\mu}$.  We then have the conditions:
\begin{equation}
l\cdot m=l \cdot \bar{m}=n\cdot m=n\cdot \bar{m}=0
\end{equation}
from orthogonality, and:
\begin{equation}
l\cdot l=n \cdot n=m\cdot m=\bar{m}\cdot \bar{m}=0
\end{equation}
since the basis vectors are null.  Finally, we impose:
\begin{equation}
l\cdot n = -m\cdot \bar{m}=1
\end{equation}
The Weyl tensor, given in terms of the Riemann and Ricci tensors,
is:
\begin{eqnarray}
C_{\mu\nu\gamma\delta}=R_{\mu\nu\gamma\delta}&-&\frac{1}{2}(\eta_{\mu
\gamma}R_{\nu\delta}-\eta_{\nu\gamma}R_{\mu\delta}-
\eta_{\mu\delta}R_{\nu\gamma}+\eta_{\nu\delta}R_{\mu\gamma})\nonumber\\
&+&\frac{1}{6}(\eta_{\mu\gamma}\eta_{\nu\delta}-\eta_{\mu\delta}\eta_{\nu\gamma})R
\label{weyl Tensor}
\end{eqnarray}
where $R_{\mu\gamma}=\eta^{\nu\delta}R_{\mu\nu\gamma\delta}$ and
$R=\eta^{\mu\nu}R_{\mu\nu}$; and the Weyl tensor satisfies the
trace-free condiation:
\begin{equation}
\eta^{\mu\delta}C_{\mu\nu\gamma\delta}=0
\end{equation}
and the cyclicity property:
\begin{equation}
C_{1234}+C_{1342}+C_{1423}=0
\end{equation}
Now, using the null tetrad, we can denote the ten independent
components of the Weyl tensor by the five complex Newman-Penrose
scalars:
\begin{eqnarray}
\Psi_0&=&-C_{\mu\nu\gamma\delta}l^{\mu}m^{\nu}l^{\gamma}m^{\delta}\nonumber\\
\Psi_1&=&-C_{\mu\nu\gamma\delta}l^{\mu}n^{\nu}l^{\gamma}m^{\delta}\nonumber\\
\Psi_2&=&-C_{\mu\nu\gamma\delta}l^{\mu}m^{\nu}\bar{m}^{\gamma}n^{\delta}\nonumber\\
\Psi_3&=&-C_{\mu\nu\gamma\delta}l^{\mu}n^{\nu}\bar{m}^{\gamma}n^{\delta}\nonumber\\
\Psi_4&=&-C_{\mu\nu\gamma\delta}n^{\mu}\bar{m}^{\nu}n^{\gamma}\bar{m}^{\delta}
\label{penrose scalars}
\end{eqnarray}
\subsection{Polarization Sum Rule}
\subsubsection{Ricci Flat Spacetime} In Ricci flat spacetime, summing
the quantum correction over the two polarizations leads to the
following polarization sum rule:
\begin{equation}
\sum_{a}\delta k(a)=0
\end{equation}
This can be proven by suming the quantum correction in
Eqn.~(\ref{quantum modified light cone correction}) over the two
polarizations,
\begin{equation}
\sum_{a}\delta
k(a)=\frac{8c_{\alpha}}{m_e^2}\sum_{a}R_{\mu\nu\gamma\delta}k^{\mu}k^{\gamma}a^{\nu}a^{\delta}
\end{equation}
In the Newman-Penrose basis, using $k=l$,
$a=\frac{\sqrt{2}}{2}(m+\bar{m})$,
$b=-\frac{i\sqrt{2}}{2}(m-\bar{m})$, and the fact that
$C_{\mu\nu\gamma\delta}=R_{\mu\nu\gamma\delta}$ for the Ricci flat
case, we have:
\begin{eqnarray}
\sum_{a}R_{\mu\nu\gamma\delta}k^{\mu}k^{\gamma}a^{\nu}a^{\delta}&=&\frac{1}{2}C_{\mu\nu\gamma\delta}l^{\mu}l^{\gamma}(m^{\nu}
+\bar{m}^{\nu})(m^{\delta}+\bar{m}^{\delta})\nonumber\\
&-&\frac{1}{2}C_{\mu\nu\gamma\delta}l^{\mu}l^{\gamma}(m^{\nu}-\bar{m}^{\nu})(m^{\delta}-\bar{m}^{\delta})\nonumber\\
&=&C_{\mu\nu\gamma\delta}l^{\mu} l^{\gamma} (m^{\nu}\bar{m}^{\delta}
+\bar{m}^{\nu}m^{\delta})
\end{eqnarray}
This particular contraction is equal to zero as it's not part of the
complex scalars in Eqns.~(\ref{penrose scalars}); hence the sum of
the two quantum corrections is zero. This implies the trajectory
(and velocity) shifts are equal and opposite.
\subsubsection{Non-Ricci Flat Spacetime} For the non-Ricci flat
spacetimes the polarization sum rule is given as:
\begin{equation}
\sum_{a}\delta
k(a)=-\frac{8\pi}{m_e^2}(2b-8c_{\alpha})T_{\mu\nu}k^{\mu}k^{\nu}
\label{polar-sum rule-non-ricci flat}
\end{equation}
where $T_{\mu\nu}$ is the energy-momentum tensor. This can be shown
by proceeding as before, but now we include the Ricci tensor and
scalar, as in Eqn.~(\ref{weyl Tensor}):
\begin{equation}
\sum_{a}R_{\mu\nu\gamma\delta}k^{\mu}k^{\gamma}a^{\nu}a^{\delta}=C_{\mu\nu\gamma\delta}l^{\mu}l^{\gamma}(m^{\nu}\bar{m}^{\delta}
+\bar{m}^{\nu}m^{\delta})-R_{\mu\gamma}l^{\mu}l^{\gamma}
\end{equation}
As before the first term on the RHS is zero, and the second term is
only dependent on the photon momentum. Then, combining this with the
Ricci term in Eqn.~(\ref{quantum modified light cone equation}) we
have:
\begin{equation}
\sum_{a}\delta
k(a)=\frac{1}{m_e^2}(2b-8c_{\alpha})R_{\mu\nu}k^{\mu}k^{\nu}
\end{equation}
Finally replacing the Ricci tensor with the energy-momentum tensor,
by using the Einstein equation
\begin{equation}
R_{\mu\nu}=-8\pi T_{\mu\nu}+\frac{1}{2}Rg_{\mu\nu}
 \label{einstein}
\end{equation}
we obtain Eqn.~(\ref{polar-sum rule-non-ricci flat}).
\subsection{Horizon Theorem}
At the event horizon, photons with momentum directed normal to the
horizon have velocity equal to $c$, i.e. the light cone remains
$k^2=0$, independent of polarization\cite{graham}.
\\
\\
This can be easily proven for the Ricci flat spacetime, using the
orthonormal vectors $k^a=(E_k,E_k,0,0)$, $a^b=(0,0,1,0)$ and
$a^b=(0,0,0,1)$.  Therefore, using these vectors in
Eqn.~(\ref{quantum modified light cone correction}), we have:
\begin{equation}
k^2=\frac{8c_{\alpha}}{m_{e}^2}R_{abcd}k^ak^ca^ba^d=0
\end{equation}
So, in Ricci flat spacetime all radially projected photon
trajectories remain unchanged.

It is also possible to prove the horizon theorem for the general
case (for Ricci and non-Ricci flat spacetimes) that the light cone
at the event horizon is unchanged.  This can be seen in the null
tetrad, so that the physical, space-like, polarization vectors
$a^{\mu}$ and $b^{\mu}$ lie parallel to the event horizon 2-surface,
while $k^{\mu}$ is the null vector normal to the surface. Then, from
Eqn.~(\ref{quantum modified light cone equation}) we have for the
two polarizations:
\begin{eqnarray}
k^2&=&\frac{2b}{m_e^2}R_{\mu\gamma}k^{\mu}k^{\gamma}+\frac{8c_{\alpha}}{m_{e}^2}R_{\mu\nu\gamma\delta}k^{\mu}k^{\gamma}a^{\nu}a^{\delta}\nonumber\\
&=&\frac{2b}{m_e^2}R_{\mu\gamma}l^{\mu}l^{\gamma}+\frac{8c_{\alpha}}{m_{e}^2}[-\frac{1}{2}R_{\mu\gamma}l^{\mu}l^{\gamma}\pm C_{\mu\nu\gamma\delta}l^{\mu}l^{\gamma}\frac{1}{2}(m^{\nu}\pm\bar{m}^{\nu})(m^{\delta}\pm\bar{m}^{\delta})]\nonumber\\
&=&\frac{1}{m_e^2}(2b-4c_{\alpha})R_{\mu\gamma}l^{\mu}l^{\gamma}\pm\frac{4c_{\alpha}}{m_e^2}C_{\mu\nu\gamma\delta}l^{\mu}l^{\gamma}\frac{1}{2}(m^{\nu}\pm\bar{m}^{\nu})(m^{\delta}\pm\bar{m}^{\delta})
\end{eqnarray}
Using Eqn.~(\ref{einstein}) and the fact that
$C_{\mu\nu\gamma\delta}l^{\mu}l^{\gamma}(m^{\nu}\bar{m}^{\delta}+\bar{m}^{\nu}m^{\delta})=0$,
we can write this as:
\begin{eqnarray}
k^2&=&-\frac{8\pi}{m_e^2}(2b-4c_{\alpha})T_{\mu\gamma}l^{\mu}l^{\gamma}\pm\frac{4c_{\alpha}}{m_e^2}C_{\mu\nu\gamma\delta}l^{\mu}l^{\gamma}(m^{\nu}m^{\delta}+\bar{m}^{\nu}\bar{m}^{\delta})
\end{eqnarray}
and in terms of the Newman-Penrose scalars this can be written as:
\begin{eqnarray}
k^2&=&-\frac{8\pi}{m_e^2}(2b-4c_{\alpha})T_{\mu\gamma}l^{\mu}l^{\gamma}\pm\frac{8c_{\alpha}}{m_e^2}\Psi_0,
\label{equation}
\end{eqnarray}
where the simplification in the last term on the RHS is possible as
$\Psi_0$ is real for Schwarzschild spacetime. In general
Eqn.~(\ref{equation}) is non zero, however, at the event horizon the
terms: $T_{\mu\gamma}l^{\mu}l^{\gamma}$ and $\Psi_0$ are zero for
stationary spacetimes\cite{graham,hawking}\footnote{Stationary
spacetimes are independent of time and may or may not be symmetric
under time reversal}.

Physically the Ricci term represents the flow of matter across the
horizon and the Weyl term represents the flow of gravitational
radiation\cite{graham}, and both are zero in classic general
relativity; and as, even with the quantum modification, the light
cone remains unchanged at the event horizon, means the event horizon
is fixed and light cannot escape from inside the black hole.

%% file: part1-chapter4.tex
\chapter{Quantum Modified Trajectories}
\label{quant-mod-trajectories-chapter} In this chapter we will
analyse how the classical null trajectories, described in
Chapter~\ref{null dynamics}, are modified in Schwarzschild
spacetime, due to quantum modifications of the equations of motion
of general relativity. Using Eqn.~(\ref{quantum modified light cone
equation}) we will calculate the quantum corrected version of
Eqn.~(\ref{du/dphi=f(u)}), which will then describe the quantum
modified motion of a null trajectory in a Schwarzschild spacetime.
Using this, and following the methods of Chapter~\ref{null
dynamics}, we will begin by studying simple critical circular orbits
at the stationary point, $u=1/3M$.  As stated in the polarization
rule, the critical orbit should be shifted up and down by equal
amounts, depending on the polarization of each photon. Also, we will
show that these modifications are only valid if the "classic" impact
parameter corresponding to the stationary orbit is adjusted,
depending on the photon's polarization, in order to compensate for
the orbit shift. This information, of the modified orbits and the
corresponding impact parameter for each polarization, will then be
used to determine the general trajectory of a (vertically or
horizontally polarized) photon coming in from infinity and tending
to one of the two shifted critical orbits. We will then go on to
show that these quantum modifications have no effect on the event
horizon, that is, when the impact parameter is accordingly adjusted
and a photon falls into the singularity the event horizon remains
fixed at $u=1/2M$.
\section{Quantum Modified Circular Orbits}
In Schwarzschild spacetime, using $k^{\nu}=dx^{\nu}/d\tau$,
Eqn.~(\ref{quantum modified light cone correction}) can be written
as:
    \begin{eqnarray}
            0&=&\dot{r}^2-(1-\frac{2M}{r})^2\dot{t}^2+r^2(1-\frac{2M}{r})\dot{\theta}^2\nonumber\\
            &+&(1-\frac{2M}{r})r^2\dot{\phi}^2+\frac{8c_{\alpha}}{m_{e}^2}(1-\frac{2M}{r})R_{\mu\nu\gamma\delta}
            k^{\mu} k^{\gamma} a^{\nu} a^{\delta}
            \label{quantum light cone}
    \end{eqnarray}
and Eqn.~(\ref{quantum modified geodesic equation ricci}) as:
    \begin{eqnarray}
            0&=&\ddot{t}+\frac{2M}{r^2}(1-\frac{2M}{r})^{-1}\dot{r}\dot{t}\nonumber\\
            &-&\frac{4c_{\alpha}}{m_e^2}\partial_t
            (R_{\mu\nu\gamma\delta}a^{\nu}
            a^{\delta}\frac{dx^{\mu}}{d\tau}\frac{dx^{\gamma}}{d\tau})
            \label{quantum trajectory t}
    \end{eqnarray}
    \begin{eqnarray}
            0&=&\ddot{r}-\frac{M}{r^2}(1-\frac{2M}{r})^{-1}\dot{r}^2-r(1-\frac{2M}{r})\dot{\phi}^2\nonumber\\
            &+&\frac{M}{r^2}(1-\frac{2M}{r})\dot{t}^2
            -\frac{4c_{\alpha}}{m_e^2}\partial_r(R_{\mu\nu\gamma\delta}a^{\nu}
            a^{\delta}\frac{dx^{\mu}}{d\tau}\frac{dx^{\gamma}}{d\tau})
             \label{quantum trajectory r}
    \end{eqnarray}
    \begin{eqnarray}
            0&=&\ddot{\phi}+\frac{2\dot{r}\dot{\phi}}{r}-\frac{4c_{\alpha}}{m_e^2}\partial_\phi
            (R_{\mu\nu\gamma\delta}a^{\nu}
            a^{\delta}\frac{dx^{\mu}}{d\tau}\frac{dx^{\gamma}}{d\tau})
             \label{quantum trajectory phi}
    \end{eqnarray}
   \begin{eqnarray}
          0&=&-\frac{4c_{\alpha}}{m_e^2}\partial_\theta
            (R_{\mu\nu\gamma\delta}a^{\nu}
            a^{\delta}\frac{dx^{\mu}}{d\tau}\frac{dx^{\gamma}}{d\tau})
             \label{quantum trajectory theta}
    \end{eqnarray}
   where $c_{\alpha}=\frac{\alpha}{360\pi}$.  Now, in order to consider quantum modified circular orbits we
require three things: (i) the Riemann tensor components, (ii) the
photon wave vectors, $k^\mu=\frac{dx^\mu}{d\tau}$, for circular
orbits, and finally (iii) the photon polarization vectors, $a^\mu$.
Due to the circular nature of the orbit the simplest basis to work
in is the orthonormal basis. In this frame the required polarization
and wave vectors, for circular orbits, are simply given as:
    \begin{equation}
        a^\mu=(0,1,0,0) \qquad \textrm{Planar Polarized}
        \label{plane polarized a}
    \end{equation}
    \begin{equation}
        a^\mu=(0,0,1,0)  \qquad \textrm{Vertically Polarized}
         \label{vertcal polarized a}
    \end{equation}
    \begin{equation}
        k^\mu=\frac{dx^\mu}{d\tau}=(E_k,0,0,E_k) \qquad \textrm{$4$-momentum
        along $\phi$}
        \label{k photon wave vector}
    \end{equation}
Now using these photon vectors, the quantum modifications in
Eqns.~(\ref{quantum light cone})-(\ref{quantum trajectory theta}),
can be expanded to give:
\begin{itemize}
    \item For the planar polarized case we have:
    \begin{eqnarray}
        R_{abcd}k^a k^c a^b a^d&=&R_{arcr} k^a k^c=E_k^2(R_{arcr} \hat{k}^a
        \hat{k}^c)\nonumber\\
        &=&E_k^2(R_{trtr}+R_{\phi r \phi r}+2R_{tr\phi r})
        \label{planar modification}
    \end{eqnarray}
    \item For the vertically polarized case we have:
    \begin{eqnarray}
        R_{abcd}k^a k^c a^b a^d&=&R_{a\theta c \theta} k^a k^c=E_k^2(R_{a \theta c \theta} \hat{k}^a
        \hat{k}^c)\nonumber\\
        &=&E_k^2(R_{t \theta t \theta
        }+R_{\phi\theta\phi\theta})
        \label{vertical modification}
    \end{eqnarray}
\end{itemize}
Using the six independent components of the Riemann tensor in the
orthonormal basis\footnote{Which we have calculated in Appendix A},
we find:
    \begin{equation}
        R_{trrt}=-R_{trtr}=-\frac{2M}{r^3} \Rightarrow
        R_{trtr}=\frac{2M}{r^3}
    \end{equation}
    \begin{equation}
        R_{\phi rr \phi}=-R_{\phi r \phi r}=-\frac{M}{r^3} \Rightarrow
        R_{\phi r \phi r}=\frac{M}{r^3}
    \end{equation}
    \begin{equation}
        R_{\theta t \theta t}=R_{t\theta t \theta}=-\frac{M}{r^3} \Rightarrow
        R_{t\theta t\theta}=-\frac{M}{r^3}
    \end{equation}
    \begin{equation}
        R_{\phi \theta \theta \phi}=-R_{\phi \theta \phi \theta}=\frac{2M}{r^3} \Rightarrow
        R_{\phi \theta \phi \theta}=-\frac{2M}{r^3}
    \end{equation}
    \begin{equation}
        R_{tr\phi r}=0
    \end{equation}
Then Eqns.~(\ref{planar modification}) and (\ref{vertical
modification}) become:
\begin{itemize}
    \item For the planar polarized case:
    \begin{equation}
        R_{abcd}k^a k^c a^b a^d=E_k^2(R_{trtr}+R_{\phi r \phi
        r}+2R_{tr\phi r})=\frac{3E_k^2M}{r^3}
        \label{planar R term}
    \end{equation}
with the relevant derivatives:
    \begin{eqnarray}
        \partial_t\frac{3E_k^2M}{r^3}&=&0 \qquad
        \partial_r\frac{3E_k^2M}{r^3}=-\frac{9E_k^2M}{r^4}\nonumber\\
        \partial_\theta\frac{3E_k^2M}{r^3}&=&0 \qquad
        \partial_\phi\frac{3E_k^2M}{r^3}=0
        \label{planar R derivative term}
    \end{eqnarray}
    \item For the vertically polarized case:
    \begin{equation}
       R_{abcd}k^a k^c a^b a^d=E_k^2(R_{t \theta t \theta
        }+R_{\phi\theta\phi\theta})=-\frac{3E_k^2M}{r^3}
        \label{vertical R term}
    \end{equation}
and the relevant derivatives:
    \begin{eqnarray}
        -\partial_t\frac{3E_k^2M}{r^3}&=&0 \qquad
        -\partial_r\frac{3E_k^2M}{r^3}=\frac{9E_k^2M}{r^4}\nonumber\\
        -\partial_\theta\frac{3E_k^2M}{r^3}&=&0 \qquad
        -\partial_\phi\frac{3E_k^2M}{r^3}=0
        \label{vertical R derivative term}
    \end{eqnarray}
\end{itemize}
From Eqns.~(\ref{quantum modified light cone correction}),
(\ref{planar R term}) and (\ref{vertical R term}) we can see that:
\begin{eqnarray}
k^2&\sim&\frac{3E_k^2M}{r^3} \qquad \textrm{For Planar
Polarization}\\
 k^2&\sim&-\frac{3E_k^2M}{r^3} \qquad \textrm{For
Vertical Polarization}
\end{eqnarray}
This implies that, as the light cone for planar polarization is
positive, it represents a photon trajectory with a speed $<c$, and
as the light cone for vertical polarization is negative, it
represents a photon trajectory with a speed $>c$.
\subsubsection{Planar Polarization}
Considering the planar polarized case first, we can use
Eqns.~(\ref{planar R term}) and (\ref{planar R derivative term}) to
rewrite Eqns.~(\ref{quantum light cone})-(\ref{quantum trajectory
phi}) as:
    \begin{eqnarray}
            0&=&\dot{r}^2-(1-\frac{2M}{r})^2\dot{t}^2+r^2(1-\frac{2M}{r})\dot{\theta}^2\nonumber\\
            &+&(1-\frac{2M}{r})r^2\dot{\phi}^2+\frac{8c_{\alpha}}{m_{e}^2}(1-\frac{2M}{r})\frac{3E_k^2M}{r^3}
            \label{quantum light cone with planar term}
    \end{eqnarray}
    \begin{eqnarray}
            0&=&\ddot{t}+\frac{2M}{r^2}(1-\frac{2M}{r})^{-1}\dot{r}\dot{t}
            \label{quantum trajectory t with planar term}
    \end{eqnarray}
    \begin{eqnarray}
            0&=&\ddot{r}-\frac{M}{r^2}(1-\frac{2M}{r})^{-1}\dot{r}^2-r(1-\frac{2M}{r})\dot{\phi}^2\nonumber\\
            &+&\frac{M}{r^2}(1-\frac{2M}{r})\dot{t}^2+\frac{4c_{\alpha}}{m_e^2}\frac{9E_k^2M}{r^4}
             \label{quantum trajectory r with planar term}
    \end{eqnarray}
    \begin{eqnarray}
            0&=&\ddot{\phi}+\frac{2\dot{r}\dot{\phi}}{r}
             \label{quantum trajectory phi with planar term}
    \end{eqnarray}
where Eqn.~(\ref{quantum trajectory theta}) becomes zero.  Now,
using the solutions of (\ref{quantum trajectory t with planar term})
and (\ref{quantum trajectory phi with planar term}), as given by
(\ref{dot-t}) and (\ref{dot-phi}), we can rewrite Eqn.~(\ref{quantum
light cone with planar term}) as a simple quantum modified
trajectory equation for circular orbits with radius $r$ and in a
plane $\theta=\frac{\pi}{2}$:\footnote{We must note that $E$ and
$E_k$ are not equal, $E$ is the energy from the classic relativistic
orbit equations, while $E_k$ is the quantum energy of the photon}
    \begin{eqnarray}
        0&=&\dot{r}^2-E^2+(1-\frac{2M}{r})\frac{J^2}{r^2}+\frac{8c_{\alpha}}{m_{e}^2}(1-\frac{2M}{r})\frac{3E_k^2M}{r^3}\nonumber\\
        \Rightarrow \qquad E^2&=&(\frac{dr}{d\tau})^2+(1-\frac{2M}{r})(\frac{24cE_k^2M}{m_e^2r^3}+\frac{J^2}{r^2})
        \label{plane equation preliminary}
    \end{eqnarray}
To make this equation more meaningful and easier to solve we make
the following transformations\footnote{Using
$J=r^2\frac{d\phi}{d\tau}$}:
    \begin{equation}
        \tau\rightarrow\phi \qquad
        \frac{dr}{d\phi}\cdot\frac{d\phi}{d\tau}=\frac{dr}{d\phi}\cdot\frac{J}{r^2}
    \end{equation}
    \begin{equation}
        r\rightarrow u=\frac{1}{r} \qquad
        \frac{du}{d\phi}\cdot\frac{
        dr}{du}=\frac{du}{d\phi}\cdot(-\frac{1}{u^2})
    \end{equation}
Therefore, Eqn.~(\ref{plane equation preliminary}) becomes:
    \begin{eqnarray}
        E^2&=&(\frac{dr}{d\phi})^2\cdot(\frac{J}{r^2})^2+(1-\frac{2M}{r})(\frac{24c_{\alpha}E_k^2M}{m_e^2r^3}+\frac{J^2}{r^2})\nonumber\\
        &=&(\frac{du}{d\phi})^2\cdot(-\frac{1}{u^2})^2\cdot(J
        u^2)^2+(1-2Mu)(\frac{24c_{\alpha}E_k^2Mu^3}{m_e^2}+J^2u^2)\nonumber\\
        &=&(\frac{du}{d\phi})^2\cdot J^2+(1-2Mu)(\frac{24c_{\alpha}E_k^2Mu^3}{m_e^2}+J^2u^2)
        \end{eqnarray}
Which can be written as:
    \begin{eqnarray}
        (\frac{du}{d\phi})^2&=&2M^2Au^4+(2M-MA)u^3-u^2+\frac{E^2}{J^2}
            \label{planar r polarization trajectory equation}
    \end{eqnarray}
where we have defined the dimensionless constant:
    \begin{equation}
        A=\frac{24c_{\alpha}E_k^2}{J^2m_e^2}=\frac{72 c_{\alpha}}{m_e^2
        D^2},
        \label{constant A}
    \end{equation}
In the last form we have used (without proof) the relation
$E_k=\sqrt{3}E$, which will be proven in Sec.~\ref{gen-vect},
Eqn.~(\ref{norm}).\footnote{$c_{\alpha}$ is dimensionless, while $D$
and $m_e$ have dimensions of length and inverse-length respectively}
\subsubsection{Vertical Polarization}
Similarly, for the vertically polarized case, using
Eqns.~(\ref{vertical R term}) and (\ref{vertical R derivative term})
in Eqns.~(\ref{quantum light cone})-(\ref{quantum trajectory theta})
to give:
    \begin{eqnarray}
            0&=&\dot{r}^2-(1-\frac{2M}{r})^2\dot{t}^2+r^2(1-\frac{2M}{r})\dot{\theta}^2\nonumber\\
            &+&(1-\frac{2M}{r})r^2\dot{\phi}^2-\frac{8c_{\alpha}}{m_{e}^2}(1-\frac{2M}{r})\frac{3E_k^2M}{r^3}
            \label{quantum light cone with vertical term}
    \end{eqnarray}
    \begin{eqnarray}
            0&=&\ddot{t}+\frac{2M}{r^2}(1-\frac{2M}{r})^{-1}\dot{r}\dot{t}
            \label{quantum trajectory t with vertical term}
    \end{eqnarray}
    \begin{eqnarray}
            0&=&\ddot{r}-\frac{M}{r^2}(1-\frac{2M}{r})^{-1}\dot{r}^2-r(1-\frac{2M}{r})\dot{\phi}^2\nonumber\\
            &+&\frac{M}{r^2}(1-\frac{2M}{r})\dot{t}^2-\frac{4c_{\alpha}}{m_e^2}\frac{9E_k^2M}{r^4}
             \label{quantum trajectory r with vertical term}
    \end{eqnarray}
    \begin{eqnarray}
            0&=&\ddot{\phi}+\frac{2\dot{r}\dot{\phi}}{r}
             \label{quantum trajectory phi with vertical term}
    \end{eqnarray}
which, in a similar way as before, gives us the equation of motion
for the vertically polarized photon:
    \begin{equation}
        (\frac{du}{d\phi})^2=-2M^2Au^4+(2M+MA)u^3-u^2+\frac{E^2}{J^2}
        \label{vertical theta polarisation trajectory equation}
    \end{equation}
\subsection{Quantum Modified Critical Orbits} Now the
general equation for the quantum modified circular orbits is:
    \begin{equation}
        (\frac{du}{d\phi})^2=\pm
        AM(2Mu-1)u^3+2Mu^3-u^2+\frac{1}{D^2}=f(u)
        \label{polarization trajectory equation}
    \end{equation}
where $+$ is for planar polarization in the $r$ direction, $-$ is
for vertical polarization in the $\theta$ direction and $D$ is the
impact parameter.  As $A\rightarrow0$ Eqn.~(\ref{polarization
trajectory equation}) tends to the classic orbit equation in general
relativity, Eqns.~(\ref{du/dphi=f(u)}).  Therefore, in order to
determine the magnitude of the quantum correction we can calculate
the order of $A$ using typical values for $m_e$, $D$ and
$c_{\alpha}$, in Eqn.~(\ref{constant A}). Using\footnote{As we were
working with $G=c=\hbar=1$, we must reintroduce these constants to
obtain the correct order of $A$} $m_e c/h\sim 10^{11}$ for the
electron mass (as it is given as inverse length),
$c_{\alpha}=\alpha/360\pi\sim 10^{-6}$, and the mass of the sun
 inserted into the critical impact
parameter: $D=3\sqrt{3} GM_{\bigodot}/c^2\sim 10^{2}$ (given in
terms of length), this then gives us\footnote{This result is also
shown in \cite{graham2}}:
\begin{eqnarray}
A=\frac{72 c_{\alpha}}{m_e^2
        D^2}\sim\frac{10^{-6}}{(10^{11} 10^{2})^2}\sim 10^{-32}
\label{order of A}
\end{eqnarray}
With the order of $A$ being so small the correction in
Eqn.~(\ref{polarization trajectory equation}) will be tiny compared
to the size of the orbit ($r=3GM/c^2\sim10^2$); therefore the
modified orbits will not differ from the classic critical orbit,
given by Eqn.~(\ref{du/dphi=f(u)}), by very much.

We will now determine the quantum modified critical circular orbits.
In order to solve Eqn.~(\ref{polarization trajectory equation}), we
can use the fact, from the polarization rule, that the modified
orbits should be shifted above and below $u=\frac{1}{3M}$ by equal
amounts depending on polarization.  So we expect a solution of the
type $u=\frac{1}{3M}\pm\delta u$; which means we can try a simple
modified solution of the form:
    \begin{equation}
        u=u_0+k u_1
        \label{trial solution}
    \end{equation}
where $u_0=\frac{1}{3M}$, $u_1$ is the quantum modification, and $k$
is a small constant depending on the quantum modification $A$.
\subsubsection{Planar Polarized Critical Orbit}
Working with Planar polarization, we can substitute the solution
(\ref{trial solution}) into the derivative $df/du$ of
Eqn.~(\ref{polarization trajectory equation}):
    \begin{equation}
        \frac{df}{du}=2AM^2u^3+3A M(2Mu-1)u^2+6Mu^2-2u=0
    \end{equation}
and as $A\ll1$ and $k\sim A$, then the only terms of relevance are
the ones first order in $k$ and $A$, everything else can be assumed
to be $\approx0$.  Therefore, we have\footnote{$u=u_0+ku_1 \qquad
u^2=u_0^2+2ku_1u_0 \qquad u^3=u_0^3+3ku_1u_0^2$}:
    \begin{displaymath}
        2AM^2u_0^3+6AM^2u_0^3-3AMu_0^2+6Mu_0^2+12Mku_1u_0-2u_0-2ku_1=0
    \end{displaymath}
    \begin{equation}
        \Rightarrow u_1=\frac{A M}{6k}u_0^2
        \label{trial solution modification term}
    \end{equation}
Now, substituting this into the trial solution (\ref{trial
solution}), we have:
    \begin{equation}
        u=u_0(1+\frac{A M}{6}u_0)
        \label{modified planar circular solution}
    \end{equation}
where $u_0=\frac{1}{3M}$ is the classic solution.  Therefore the
classic orbit $u_0$ is modified by $\frac{M}{6}u_0$ to first order
in $A$. Also, with this orbit modification we require an associated,
modified, impact parameter, which should take the form:
    \begin{equation}
        \frac{1}{D^2}=\frac{1}{D_0^2}+\beta
        \label{trial impact solution}
    \end{equation}
where $1/D_0^2=1/27M^2$.  The modified impact parameter can be found
by substituting (\ref{modified planar circular solution}) and
(\ref{trial impact solution}) into (\ref{polarization trajectory
equation}) and solving for $\beta$. Doing so, we find\footnote{we,
again, work to first order in A: $u=u_0+A\frac{M}{6}u_0^2$,
$u^2=u_0^2+A\frac{M}{3}u_0^3$, $u^3=u_0^3+A\frac{M}{2}u_0^4$ and
$u^4=u_0^4+A\frac{2M}{3}u_0^5$}:
    \begin{eqnarray}
        (\frac{du}{d\phi})^2&=&2M^2Au^4-AMu^3+2Mu^3-u^2+\frac{1}{D_0^2}+\beta=0\nonumber\\
        &=&2M^2Au_0^4-AMu_0^3+2Mu_0^3+M^2Au_0^4-u_0^2\nonumber\\
        &-&\frac{1}{3}AMu_0^3+\frac{1}{D_0^2}+\beta
    \end{eqnarray}
We have $2Mu_0^3-u_0^2+\frac{E^2}{J^2}=0$, as this forms the classic
equation of motion for circular orbits.  Therefore,
    \begin{displaymath}
        2M^2Au_0^2-AMu_0^3+M^2Au_0^4-\frac{1}{3}AMu_0^3+\beta=0
    \end{displaymath}
Now, as $u_0=\frac{1}{3M}$, we have:
    \begin{equation}
        \beta=\frac{A M}{3}u_0^3
    \end{equation}
Therefore the modified impact parameter, for planar polarization,
is:
    \begin{equation}
        \frac{1}{D^2}=\frac{1}{D_0^2}+\frac{A M}{3}u_0^3
    \end{equation}
Then, substituting for $D_0$, we have:
    \begin{eqnarray}
        \frac{1}{D^2}&=&\frac{1}{3(3M)^2}+\frac{AM}{3}u_0^3=\frac{u_0^2}{3}+\frac{AM}{3}u_0^3\nonumber\\
        &=&\frac{u_0^2}{3}(1+AMu_0)
        \label{modified planar impact perimeter}
    \end{eqnarray}
\subsubsection{Vertical Polarized Critical Orbit}
Doing the same for the vertically polarized photon, i.e. by using:
    \begin{equation}
       \frac{df}{du}=-2AM^2u^3-3AM(2Mu-1)u^2+6Mu^2-2u=0
    \end{equation}
and substituting the trial solution (\ref{trial solution}) we find:
    \begin{displaymath}
        -2AM^2u_0^3-6AM^2u_0^3+3AMu_0^2+6Mu_0^2+12Mku_1u_0-2u_0-2ku_1=0
    \end{displaymath}
    \begin{equation}
        \Rightarrow u_1=-\frac{AM}{6k}u_0^2
    \end{equation}
    \begin{equation}
        u=u_0(1-\frac{A M}{6}u_0)
        \label{modified vertical circular solution}
    \end{equation}
which is equal, but opposite in sign, to (\ref{trial solution
modification term}), as is expected from the polarization rule.
Also, as before, the relevant impact parameter is given by
substituting (\ref{modified vertical circular solution}) and
(\ref{trial impact solution}) into the negative equation of
(\ref{polarization trajectory equation}) and working to first order
in $A$.
    \begin{eqnarray}
        \frac{du}{d\phi}&=&-2M^2Au^4+AMu^3+2Mu^3-u^2+\frac{1}{D_0^2}+\beta=0\nonumber\\
        &=&-2M^2Au_0^4+AMu_0^3+2Mu_0^3-MAu_0^4-u_0^2\nonumber\\
        &+&\frac{1}{3}AMu_0^3+\frac{1}{D_0^2}+\beta
    \end{eqnarray}
Eliminating terms and rearranging, as before, we find:
    \begin{equation}
        \beta=-\frac{AM}{3}u_0^3
    \end{equation}
Therefore the modified impact perimeter, for vertical polarization,
is:
    \begin{eqnarray}
        \frac{1}{D^2}&=&\frac{1}{3(3M)^2}-\frac{AM}{3}u_0^3=\frac{u_0^2}{3}-\frac{AM}{3}u_0^3\nonumber\\
        &=&\frac{u_0^2}{3}(1-AMu_0)
            \label{modified vertical impact perimeter}
    \end{eqnarray}
\subsubsection{The Modified Orbits}
\begin{figure}
    \begin{center}
        \includegraphics[width=0.8\textwidth]{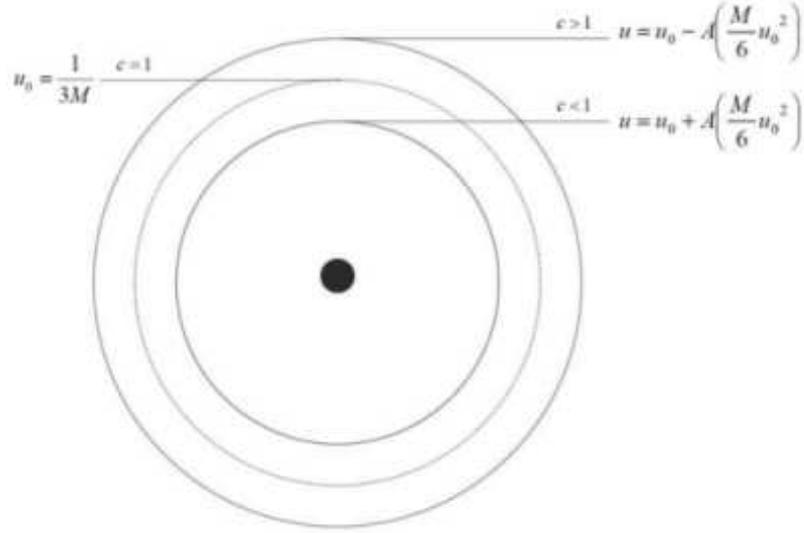}\\
        \caption{Orbit shifts about the classic critical orbit $u_0=1/3M$, with corresponding shifts in the speed of light (Not to scale)}.
        \label{orbs}
        \end{center}
\end{figure}

We now have the circular orbit solutions for Eqn.~(\ref{polarization
trajectory equation}) and the relevant impact parameters:
\begin{itemize}
    \item Planar $(r)$ polarization solution (c<1)
    \begin{equation}
        u=u_0+A(\frac{M}{6}u_0^2) \qquad \frac{1}{D^2}=\frac{u_0^2}{3}(1+A Mu_0)
        \label{modified planar solution}
    \end{equation}
    \item Vertical($\theta$) polarization solution (c>1)
    \begin{equation}
        u=u_0-A(\frac{M}{6}u_0^2) \qquad \frac{1}{D^2}=\frac{u_0^2}{3}(1-A Mu_0)
            \label{modified vertical solution}
    \end{equation}
\end{itemize}
which are displayed in Fig.~\ref{orbs} We can note that as the
constant $A$ is of the order $10^{-32}$ these modifications are
extremely small.

\section{Quantum Modified General Geodesics}
Using the solutions for the critical circular orbits and the
associated impact parameters we can now study how the general photon
trajectories are modified due to quantum corrections. In the classic
case when a photon comes in from infinity, with the critical impact
parameter, it tends to the critical orbit, as was shown in
Fig.~\ref{u vs phi with D=critical}; and if we slightly decrease the
impact parameter the photon spirals into the singularity,
Fig.~\ref{u vs phi with D=critical-0.1}.  We will now construct a
general quantum modified equation of motion and determine how the
geodesics change, depending on polarization, as they tend to the
critical orbits.
\subsection{General Vectors}
\label{gen-vect} From the quantum modification term in
Eqn.~(\ref{quantum light cone}) we can see that in order to
construct a general quantum modified geodesic equation we require
general photon polarization and wave vectors.  The wave vector,
$k^{\mu}$, can no longer be represented by a simple constant vector
pointing in the $\phi$ direction, as given in the orthonormal frame.
And, even though the vertical polarization will remain a constant,
as before, the planar polarization will now be a more general
vector, constantly changing as the photon moves through the plane.
\subsubsection{General Wave Vectors}
Our general wave vector will be of the form:
    \begin{equation}
        k^\mu=(\frac{dt}{d\tau},\frac{dr}{d\tau},\frac{d\theta}{d\tau},\frac{d\phi}{d\tau})
        \label{general wave}
    \end{equation}
which respects the light cone condition:
    \begin{equation}
        0=g_{\mu\nu}k^{\nu} k^{\mu}=F\dot{t}^2-F^{-1}\dot{r}^2-r^2\dot{\phi}^2
        \label{light cone for quantum general solution}
    \end{equation}
where $g_{\mu\nu}$ is the Schwarzschild metric.  Using previous
results, of Eqns~(\ref{dot-t}) and (\ref{dot-phi}), and  the fact
that we are working in the plane, $\theta=\frac{\pi}{2}$, we can
represent three of the wave vector components as:
    \begin{equation}
        \frac{dt}{d\tau}=EF^{-1} \qquad
        \frac{d\phi}{d\tau}=\frac{J}{r^2} \qquad
        \frac{d\theta}{d\tau}=0
        \label{wave vector components}
    \end{equation}
where we have defined:
    \begin{equation}
        F=(1-\frac{2M}{r})
        \label{F=1-2m/r}
    \end{equation}
Now using Eqn.~(\ref{light cone for quantum general solution}) we
can write the final component as:
        \begin{eqnarray}
            (\frac{dr}{d\tau})^2&=&F^2(\frac{dt}{d\tau})^2-r^2F(\frac{d\phi}{d\tau})^2\nonumber\\
            \frac{dr}{d\tau}&=&\sqrt{E^2-\frac{J^2F}{r^2}}=E(1-\frac{D^2F}{r^2})^{\frac{1}{2}}
            \label{dr/dtau}
        \end{eqnarray}
where we have used $D^2=\frac{J^2}{E^2}$.  Now the general wave
vector can be written as:
    \begin{equation}
        k^\mu=(EF^{-1},E(1-\frac{D^2F}{r^2})^{\frac{1}{2}},0,\frac{J}{r^2})
    \end{equation}
Also:
    \begin{eqnarray}
        k_{\mu}&=&g_{\mu\nu}k^\nu\nonumber\\
        k_{\mu}&=&(E,-F^{-1}E(1-\frac{D^2F}{r^2})^{\frac{1}{2}},0,-J)
    \end{eqnarray}
We now have:
    \begin{displaymath}
        k_{\mu}
        k^{\mu}=E^2F^{-1}-F^{-1}E^2(1-\frac{D^2F}{r^2})-\frac{J^2}{r^2}=0
    \end{displaymath}
as required. Therefore, eliminating $E$, we have the general wave
vectors:
    \begin{eqnarray}
           k^{\mu}&=&\frac{J}{D}(F^{-1},(1-\frac{D^2F}{r^2})^{\frac{1}{2}},0,\frac{D}{r^2})\\
            \label{k^mu general vector}
           k_{\mu}&=&\frac{J}{D}(1,-F^{-1}(1-\frac{D^2F}{r^2})^{\frac{1}{2}},0,-D)
            \label{k_mu general vector}
    \end{eqnarray}
We now need to normalize these vectors so, as they come in from
infinity and tend to the critical circular orbit, they correspond to
the critical orbit vector (\ref{k photon wave vector}). However, as
(\ref{k photon wave vector}) is given in the orthonormal basis, and
we are now working in the coordinate frame, we must use the tetrad
(\ref{inverse tetrad}), given in Appendix B, to transform (\ref{k
photon wave vector}) to its coordinate frame equivalent. Therefore,
(\ref{k photon wave vector}) in the coordinate basis is given as:
    \begin{equation}
         k^{\mu}=(e^{-1})^{\mu}_{a}k^a=(e^{-1})^{\mu}_{a}\left(%
        \begin{array}{c}
         E_k \\
         0 \\
         0 \\
         E_k\\
        \end{array}%
        \right)=E_k\left(%
        \begin{array}{c}
        F^{-\frac{1}{2}} \\
        0 \\
        0 \\
        \frac{1}{r} \\
        \end{array}%
        \right)
        \label{circular orbit transformation}
    \end{equation}
This now corresponds to a wave vector for a circular orbit in the
coordinate frame.  To represent the critical orbit we simply
substitute for $r=3M$, in which case
$F=1-\frac{2M}{r}\rightarrow\frac{1}{3}$, thus (\ref{circular orbit
transformation}) becomes:
    \begin{equation}
        k^{\mu}=E_k(\sqrt{3},0,0,\frac{1}{3M})
        \label{critical orbit in coordinate frame}
    \end{equation}
Now, if we evaluate (\ref{k^mu general vector}) with $r=3M$ and
$D=3\sqrt{3}M$ we have:
    \begin{equation}
        k^\mu=\frac{J}{D}\sqrt{3}(\sqrt{3},0,0,\frac{1}{3M})
        \label{general wavevector at critical point}
    \end{equation}
Therefore, (\ref{general wavevector at critical point}) is similar
to (\ref{critical orbit in coordinate frame}) up to a constant of
normalization, given as:
\begin{equation}
E_k=\frac{J}{D}\sqrt{3}=E\sqrt{3} \label{norm}
\end{equation}
Then, the normalized form of (\ref{k^mu general vector}) is given by
using (\ref{norm}):
    \begin{equation}
        k^{\mu}=\frac{E_k}{\sqrt{3}}(F^{-1},(1-\frac{D^2F}{r^2})^{\frac{1}{2}},0,\frac{D}{r^2})
        \label{normalized general wavevector}
    \end{equation}
\subsubsection{General Polarization Vectors}
\label{polarization vectors} Now we need to construct the, planar
and vertical, polarization vectors, $a^{\mu}$, of the photon; which
must be spacelike normalized as:
    \begin{equation}
        a_{\mu}a^{\mu}=-1 \qquad a_{\mu} k^{\mu}=0
        \label{polarization wave vector conditions}
    \end{equation}
As before, as we are working in a plane, the vertical polarization,
$ a^{\mu}_1 $, will be a constant, and can simply be written as:
    \begin{equation}
        a_{1}^{\mu}=(0,0,1,0)
        \label{vertical polarization general vector}
    \end{equation}
To normalize this we do as follows:
    \begin{equation}
        a_{1\mu}=g_{\mu\nu}a^{\nu}_1=-r^2(0,0,1,0) \qquad
        \Rightarrow \qquad a_{1}^{\mu} a_{1\mu}=-r^2
    \end{equation}
Therefore, the normalized vertical polarization vector is given as:
    \begin{eqnarray}
        a_1^{\mu}&=&(0,0,\frac{1}{r},0)\nonumber\\
        a_{1\mu}&=&-r^2(0,0,\frac{1}{r},0)
        \label{vertical polarization normailized general vector}
    \end{eqnarray}
These now satisfy both the conditions in (\ref{polarization wave
vector conditions}).  The planar polarized vector, $ a^{\mu}_2 $, is
given in the plane of $r$ and $\phi$:
    \begin{equation}
        a^{\mu}_2=(0,A,0,B)
        \label{planar vector}
    \end{equation}
Now, using the two conditions in (\ref{polarization wave vector
conditions}) we can determine $A$ and $B$.  From:
    \begin{equation}
        a_{2\mu}=g_{\mu\nu}a_2^{\nu}=(0,-AF^{-1},0,-r^2B)
        \label{a_mu plane polarization}
    \end{equation}
and the vector (\ref{normalized general wavevector}) we have:
    \begin{equation}
        k^{\mu}a_{2\mu}=(1-\frac{D^2F}{r^2})^{\frac{1}{2}}(AF^{-1})+(\frac{D}{r^2})(r^2B)=0
        \label{wave and polarization}
    \end{equation}
    \begin{equation}
        a^{2\mu}a_{2\mu}=A^2F^{-1}+B^2r^2=-1
    \end{equation}
Solving these for $A$ and $B$:
    \begin{equation}
        B=-\frac{F^{-1}}{D}(1-\frac{D^2F}{r^2})^{\frac{1}{2}}A
    \end{equation}
    \begin{equation}
        A^2=\frac{F}{(1+\frac{r^2F^{-1}}{D^2}(1-\frac{D^2F}{r^2}))}
    \end{equation}
    Therefore, we have:
    \begin{equation}
        A=\frac{DF}{r} \qquad
        B=-\frac{1}{r}(1-\frac{D^2F}{r^2})^{\frac{1}{2}}
    \end{equation}
and the planar polarization vector becomes:
    \begin{equation}
        a_{2}^{\mu}=(0,\frac{DF}{r},0,-\frac{1}{r}(1-\frac{D^2F}{r^2})^{\frac{1}{2}})
    \end{equation}
We now have the required polarization vectors:
    \begin{eqnarray}
        a_1^{\mu}&=&\frac{1}{r}(0,0,1,0)
        \label{final theta polarization}\\
        a_{1\mu}&=&-r^2(0,0,\frac{1}{r},0)
        \label{final theta polarization lower}\\
        a_2^{\mu}&=&\frac{1}{r}(0,DF,0,-(1-\frac{D^2F}{r^2})^{\frac{1}{2}})
        \label{final r-phi polarization}\\
        a_{2\mu}&=&(0,-\frac{D}{r},0,(1-\frac{D^2F}{r^2})^{\frac{1}{2}})
        \label{final r-phi polarization lower}
    \end{eqnarray}
where subscript 1 and 2 are vertical and planar polarizations
respectively. These now satisfy the conditions in (\ref{polarization
wave vector conditions}) with the wave vector (\ref{normalized
general wavevector}).
\newpage
\subsection{Quantum Modification}
\label{quantum mod section} Having derived the general polarization
and wave vectors, we can see from Eqn.~(\ref{quantum modified light
cone correction}) that the quantum modification given by:
    \begin{equation}
        \delta k(a)=\frac{8c_{\alpha}}{m_e^2}R_{abcd}k^ak^ca^ba^d
        \label{quantum correction}
    \end{equation}
also requires the Riemann tensor components in the coordinate frame.
In Appendix A we have calculated the required components as:
    \begin{eqnarray}
        R'_{trrt}&=&-\frac{2M}{r^3} \qquad  R'_{\theta r r
        \theta}=-\frac{MF^{-1}}{r} \qquad R'_{\phi rr
        \phi}=-\frac{MF^{-1}}{r}\nonumber\\
         R'_{\phi \theta \theta
        \phi}&=&2Mr \qquad  R'_{\theta t \theta t}=-\frac{M F}{r}
        \qquad R'_{\phi t \phi t}=-\frac{M F}{r}
        \label{coordinate frame Riemann components}
    \end{eqnarray}
Using this information we will now determine the form of the general
quantum modification; and from the polarization rule, this quantum
correction should satisfy the condition: $\delta k_1^2=-\delta
k_2^2$ for the two polarizations.

For vertical $(\theta)$ polarization we have, by using
(\ref{normalized general wavevector}) and (\ref{final theta
polarization}) in (\ref{quantum correction}):
    \begin{eqnarray}
       R_{abcd}k^ak^ca_1^ba_1^d & = & R_{a\theta c\theta} \frac{1}{r^2}
        k^ak^c
           =  R_{t\theta t \theta}\frac{1}{r^2}k^tk^t+R_{r\theta
         r\theta}\frac{1}{r^2}k^rk^r+R_{\phi\theta\phi\theta}\frac{1}{r^2}k^{\phi}k^{\phi}\nonumber\\
         & =&  -\frac{M F}{r} (\frac{1}{r^2}) (\frac{F^{-1} E_k}{\sqrt{3}})(\frac{F^{-1}
         E_k}{\sqrt{3}})\nonumber\\
         &+& \frac{M F^{-1}}{r} (\frac{1}{r^2}) (\frac{\sqrt{1-\frac{D^2F}{r^2}}
         E_k}{\sqrt{3}})(\frac{\sqrt{1-\frac{D^2F}{r^2}}E_k}{\sqrt{3}})\nonumber\\
         &-& 2Mr(\frac{1}{r^2})(\frac{DE_k}{r^2\sqrt{3}})(\frac{DE_k}{r^2\sqrt{3}})\nonumber\\
         &=&-\frac{ME_k^2F^{-1}}{3r^3}+\frac{ME_k^2F^{-1}(1-\frac{D^2F}{r^2})}{3r^3}-\frac{2ME_k^2D^2}{3r^5}\nonumber\\
         &=&-\frac{ME_k^2D^2}{r^5}
         \label{correction vertical}
    \end{eqnarray}
Similarly, for planar ($r-\phi$ plane) polarization we have, by
using (\ref{normalized general wavevector}) and (\ref{final r-phi
polarization}) in (\ref{quantum correction}):
    \begin{eqnarray}
         R_{abcd}k^ak^ca_2^ba_2^d & = &
         R_{arcr}k^{a}k^{c}a_2^ra_2^r+R_{arc\phi}k^{a}k^{c}a_2^ra_2^{\phi}
         +R_{a\phi c r}k^{a}k^{c}a_2^{\phi}a_2^{r}\nonumber\\
         &+&R_{a\phi c\phi}k^{a}k^{c}a_2^{\phi}a_2^{\phi}\nonumber\\
         &=&  R_{trtr}k^{t}k^{t}a_2^ra_2^r+R_{\phi r \phi
         r}k^{\phi}k^{\phi}a_2^ra_2^{r}+R_{\phi r r
         \phi}k^{\phi}k^{r}a_2^{r}a_2^{\phi}\nonumber\\
         &+&R_{r\phi\phi r}k^{r}k^{\phi}a_2^{\phi}a_2^{r}
         +R_{t \phi t \phi}k^{t}k^{t}a_2^{\phi}a_2^{\phi}
         +R_{r\phi r \phi}k^{r}k^{r}a_2^{\phi}a_2^{\phi}\nonumber\\
         &=&\frac{2M}{r^3}(\frac{F^{-1}E_k}{\sqrt{3}})^2(\frac{D
         F}{r})^2+\frac{M F^{-1}}{r}(\frac{D E_k}{\sqrt{3}
         r^2})^2(\frac{D F}{r})^2\nonumber\\
         &+&\frac{M F^{-1}}{r}(\frac{D E_k}{\sqrt{3}
         r^2})(\frac{E_k}{\sqrt{3}}\sqrt{1-\frac{D^2F}{r^2}})(\frac{D
         F}{r})(\frac{\sqrt{1-\frac{D^2F}{r^2}}}{r})\nonumber\\
         &+&\frac{M F^{-1}}{r}(\frac{E_k\sqrt{1-\frac{D^2F}{r^2}}}{\sqrt{3}})(\frac{D E_k}{\sqrt{3}
         r^2})(\frac{\sqrt{1-\frac{D^2F}{r^2}}}{r})(\frac{D
         F}{r})\nonumber\\
         &-&\frac{M
         F}{r}(\frac{F^{-1}E_k}{\sqrt{3}})^2(\frac{\sqrt{1-\frac{D^2F}{r^2}}}{r})^2\nonumber\\
         &+&\frac{M F^{-1}}{r}(\frac{E_k\sqrt{1-\frac{D^2F}{r^2}}}{\sqrt{3}})^2
         (\frac{\sqrt{1-\frac{D^2F}{r^2}}}{r})^2\nonumber\\
         &=&\frac{M E_k^2 D^2}{r^5}
         \label{correction planar}
    \end{eqnarray}
Therefore, the quantum modifications, $k^2=\delta k (a)$, for the
two polarizations $a_1$ and $a_2$ (vertical and planar respectively)
are:
    \begin{eqnarray}
        \delta k(a_1)&=&-(\frac{8c_{\alpha}}{m_e^2})\frac{ME_k^2 D^2}{r^5} \qquad \textrm{Vertical}\\
        \delta k(a_2)&=&(\frac{8c_{\alpha}}{m_e^2})\frac{ME_k^2
        D^2}{r^5} \qquad \textrm{Planar}
    \end{eqnarray}
where $-\delta k(a_1)=\delta k(a_2)$, as required.  Now, using
$k^{\mu}k_{\mu}-\delta k(a)$=0 we can write the general equations of
motion for the two polarizations:
    \begin{displaymath}
        k^{\mu}k_{\mu}\pm(\frac{8c_{\alpha}}{m_e^2})\frac{ME_k^2 D^2}{r^5}=0
    \end{displaymath}
    \begin{displaymath}
        (\frac{dr}{d\tau})^2=F^2(\frac{dt}{d\tau})^2-r^2F(\frac{d\phi}{d\tau})^2\pm F(\frac{8c_{\alpha}}{m_e^2})\frac{ME_k^2 D^2}{r^5}
    \end{displaymath}
As before, substituting for $\dot{t}$, $\dot{\phi}$ and transforming
$r\rightarrow\frac{1}{u}$, we have:
    \begin{equation}
        (\frac{du}{d\phi})^2=\frac{1}{D^2}-Fu^2\pm (\frac{D^2 M A}{3})Fu^5
        \label{genreal equation of motion dependent on phi}
    \end{equation}
We can also write $u$ as a function of time:
    \begin{equation}
        (\frac{du}{dt})^2=(Du^2F)^2(\frac{1}{D^2}-Fu^2\pm (\frac{D^2 M A}{3})Fu^5)
         \label{genreal equation of motion dependent on t}
    \end{equation}
where we have used the substitution for $A$ given in equation
(\ref{constant A}).  Now, the Eqns.~\ref{genreal equation of motion
dependent on phi} and \ref{genreal equation of motion dependent on
t} are the general equations of motion, $+$ for vertical ($\theta$)
polarization and $-$ for planar ($r$-$\phi$) polarization.  In these
equations we not only use the impact parameter of the form $1/D^2$
but also $D^2$, therefore the parameters for the two polarizations
are given as:
\begin{equation}
\frac{1}{D^2}=\frac{u_0^2}{3}(1\pm A Mu_0) \rightarrow
    D=\frac{\sqrt{3}}{u_0} \mp\frac{\sqrt{3}A M}{2}+\mathcal{O}(A)
\label{new impact}
\end{equation}
\subsection{General Trajectory to the Critical Orbit.}
Now that we have the general orbit equation (\ref{genreal equation
of motion dependent on phi}) we can first test whether, for the
critical impact perimeter $D$, the equation
$\frac{du}{d\phi}\rightarrow0$ for first order in $A$. Substituting
the critical orbits and the impact parameters given in
(\ref{modified planar solution}),(\ref{modified vertical solution})
and (\ref{new impact}) into (\ref{genreal equation of motion
dependent on phi}) and expanding up to first order in $A$ we have:
For planar polarization.
    \begin{eqnarray}
        (\frac{du}{d\phi})^2&=&\frac{u_0^2}{3}(1+A Mu_0)-[1-2M(u_0(1+\frac{A M}{6}u_0))[(u_0(1+\frac{A M}{6}u_0)]^2\nonumber\\
        &-&(\frac{\frac{3}{u_0^2(1+A M
        u_0)}A M}{3})[1-2M[u_0(1+\frac{A M}{6}u_0)]][u_0(1+\frac{A M}{6}u_0)]^5\nonumber\\
        &\rightarrow&0
    \end{eqnarray}
and similarly for vertical polarization.
       \begin{eqnarray}
        (\frac{du}{d\phi})^2&=&\frac{u_0^2}{3}(1-A Mu_0)-[1-2M[u_0(1-\frac{A M}{6}u_0)]][u_0(1-\frac{A M}{6}u_0)]^2\nonumber\\
        &-&(\frac{\frac{3}{u_0^2(1-A M
        u_0)}A M}{3})[1-2[(u_0(1-\frac{A M}{6}u_0)]][u_0(1-\frac{A M}{6}u_0)]^5\nonumber\\
        &\rightarrow&0
    \end{eqnarray}
Expanding these and eliminating all terms of order $A^2$ and higher,
we find that the right hand sides go to zero, as required.  Thus for
the appropriate impact parameters these equations behave as they
should.

\begin{figure}
    \begin{center}
        \includegraphics[width=0.9\textwidth]{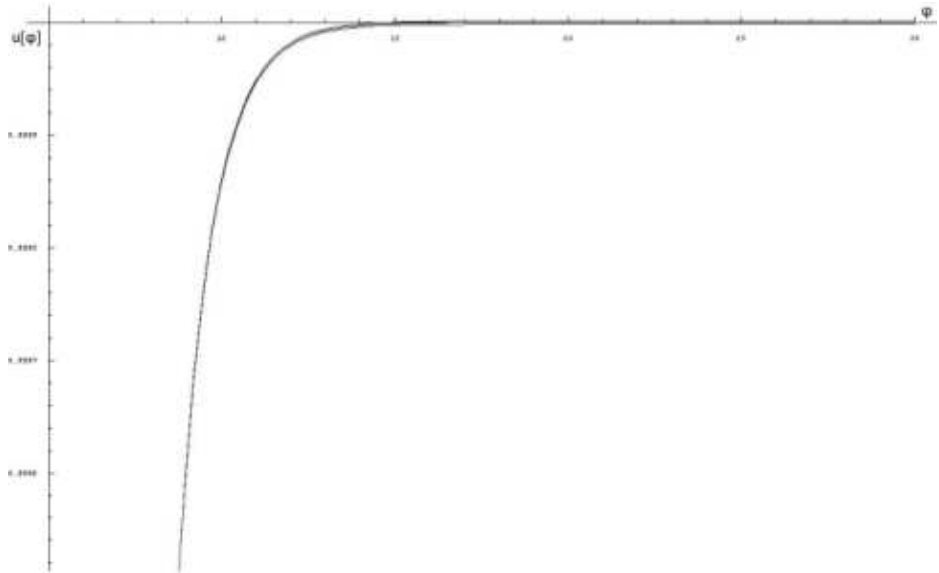}\\
        \caption{The quantum modified critical orbits follow classic
        type paths.  We have used a factor $A=0.00009$ in order to make the quantum corrections more visible,
        where a real value should be of order $\sim10^{-32}$, Eqn.~\ref{order of A}.
        The splitting of the orbits can be clearly seen in
        the Fig.~\ref{general critical trajectories large}, given below.}
        \label{general critical trajectories small}
    \end{center}
\end{figure}

The next step is to solve equation (\ref{genreal equation of motion
dependent on phi}) for the two polarizations.  This is most simply
done using numerical methods in Mathematica.  As a guide, we know
our solution will be of the form $u=u_0+k u_1$, where $u_0$ will be
the classical solution (\ref{u=critical sol}), $u_1$ will be a small
modification that pushes the critical orbit up or down depending on
photon polarization, and $k$ will be some constant that is first
order in $A$, i.e. of the form $k=A s$, where $s$ will be some
number given by the boundary condition: for $\phi\rightarrow\infty$
then $u(\phi)\rightarrow u_0(1\pm\frac{A}{6}u_0)$. If we substitute
for the critical impact parameter $D$ from (\ref{modified planar
solution}) and (\ref{modified vertical solution}) depending on the
polarization, and then transform to $u\rightarrow u_0+A s u_1$, and
expand to first order in $A$, we have non-linear first order
differential equations in $u_0$, $u_1$ and $\phi$.
\begin{itemize}
    \item For planar polarization:
    \begin{eqnarray}
        \frac{du_1(\phi)}{d\phi}&=&\frac{}{6s\sqrt{\frac{1}{27M^2}-u_0(\phi)^2+2Mu_0(\phi)^3}}
        [\frac{1}{27M^2}-27M^3u_0(\phi)^5\nonumber\\
        &+&54M^4u_0(\phi)^6- 6s u_0(\phi)u_1(\phi)+18 s Mu_0(\phi)^2u_1(\phi)]
        \label{u1(phi) equation for plane polarization}
    \end{eqnarray}
    \item For vertical polarization:
    \begin{eqnarray}
   \frac{du_1(\phi)}{d\phi}&=&\frac{1}{6s\sqrt{\frac{1}{27M^2}-u_0(\phi)^2+2Mu_0(\phi)^3}}
        [-\frac{1}{27M^2}+27M^3u_0(\phi)^5\nonumber\\
        &-&54M^4u_0(\phi)^6-6 s u_0(\phi)u_1(\phi)+18 s Mu_0(\phi)^2u_1(\phi)]
        \label{u1(phi) equation for theta polarization}
    \end{eqnarray}
\end{itemize}
These can be solved in one of two ways, (i) is to substitute the
classic solution (\ref{u=critical sol}) for $u_0$ and solve
analytically, (ii) is to solve $\frac{du_1}{d\phi}$ and the classic
equation for $\frac{du_0}{d\phi}$ simultaneously using numerical
techniques. We attempted to use method (i) to derive an analytic
solution for $u_1(\phi)$, however, due to the complex nature of the
equation we proceeded to use method (ii), that is solving by
numerical methods. To do this we set the constant $s=1$, and then
when the numerical values of $u_1(\phi)$ were determined we could
determine the constant $s$ so that $u(\phi)$ coincided with the
modified circular orbits given in (\ref{modified planar solution})
and (\ref{modified vertical solution}).  This technique was used for
reasons of convenience, because solving (\ref{u1(phi) equation for
plane polarization}) and (\ref{u1(phi) equation for theta
polarization}) for various values of $s$ would be time consuming as
each numerical calculation takes a significant amount of time;
therefore solving them once and then scaling the solution is a more
convenient method. The results of the equations were plotted as:
    \begin{equation}
        u(\phi)=u_0(\phi)+A s u_1(\phi)
        \label{general test equation}
    \end{equation}
where the constant $s$ was picked to satisfy the condition:
$\phi\rightarrow\infty$ $\Rightarrow $   $u(\phi)\rightarrow
u_0(1\pm\frac{A}{6}u_0)$ i.e. the trajectories tend to the critical
orbit, depending on polarization. In this way the constant $s$ was
determined to be $s=1/3$, which was tested for various values of
$M$. We have plotted the results of the numerical calculation in
Figs.~\ref{general critical trajectories small} and \ref{general
critical trajectories large}.  In Fig.~\ref{general critical
trajectories small}, you can see that the general critical orbits
follow a classic style path, however the orbit splitting is not
clearly visible.  But in Fig.~\ref{general critical trajectories
large} we have plotted a closer view of the critical orbits, and
here the splitting is highly visible.  It can be seen that the
general trajectories for the planar and vertically polarized photons
tend to the relevant critical orbits.
\begin{figure}
    \begin{center}
        \includegraphics[width=0.9\textwidth]{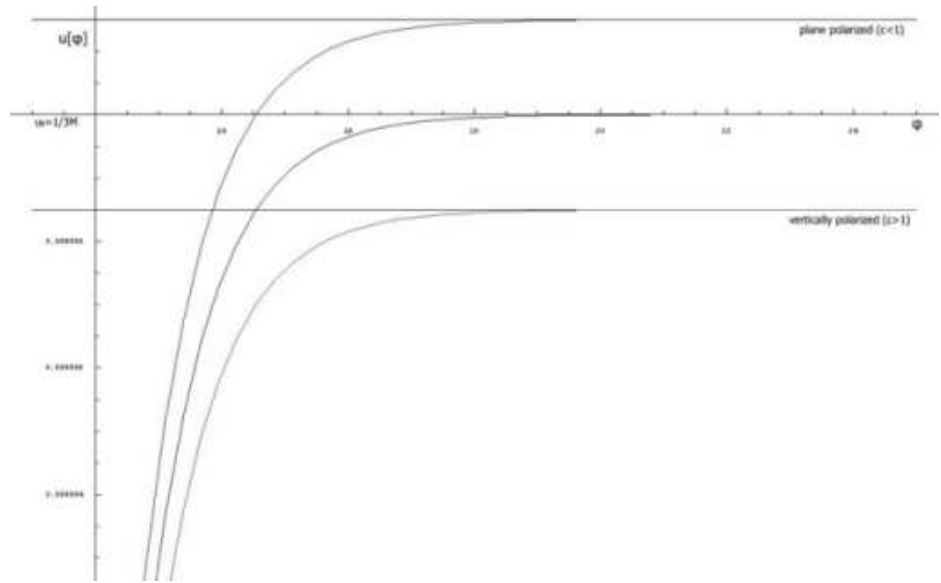}\\
        \caption{The quantum modified trajectories show clear splitting as they tend to the critical points.
        We used the constants $M=\frac{1}{3}$ and $A=0.00009$.}\label{general critical trajectories large}
    \end{center}
\end{figure}
\newpage
\section{Quantum Modification and the Event Horizon}
In Chapter.~\ref{null dynamics} it was shown that when we decreased
the impact parameter from the critical value ($D=3\sqrt{3}M$ to
$D-1/10$) the photon trajectory, given as $u(\phi)$, spiraled into
the singularity, Fig.~\ref{u vs phi with D=critical-0.1}; and when
we represent the trajectory as a function of coordinate time,
$u(t)$, it tends to the event horizon, $u=1/2M$. From the horizon
theorem it was seen that quantum modifications have no effect on
photon velocities directed normal to the event horizon.  So this
implies when a photon tends to the event horizon at an angle, e.g.
with an impact parameter $D=D_{critical}-1/10$, then the component
of velocity normal to the horizon should be unchanged, while the
component parallel to it is modified according to the quantum
correction; this modification should then result in a shift of the
photon trajectory, but the horizon should remain fixed. In order to
test this we used Eqn.~(\ref{genreal equation of motion dependent on
phi}) with an impact parameter $D=D_{critical}-1/10$ to show that
the photon trajectories still fall into the singularity. We then
used Eqn.~(\ref{genreal equation of motion dependent on t}), with
the new impact parameter, to study the behavior of the trajectories
around the event horizon.
\subsection{Trajectories to the singularity}
\begin{figure}
     \begin{center}
        \includegraphics[width=0.9\textwidth]{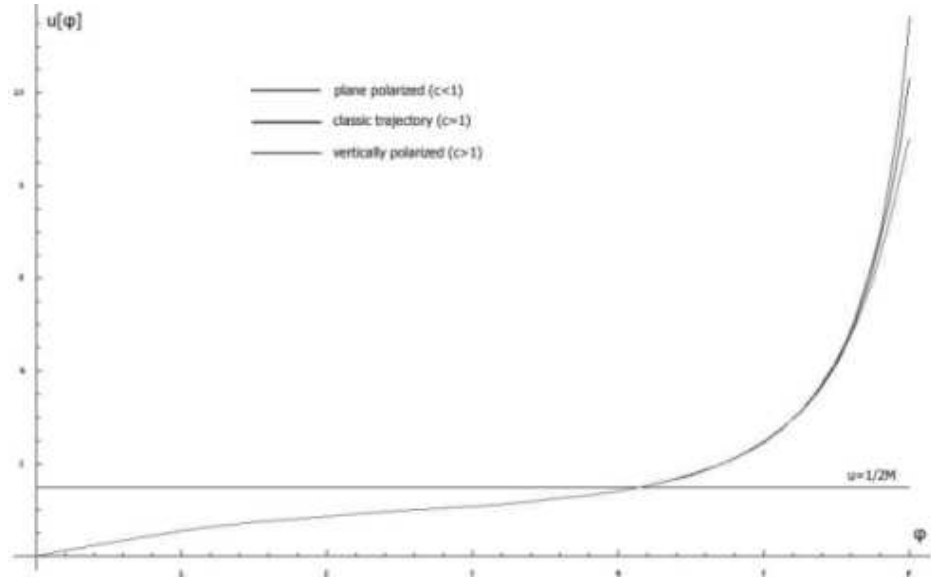}\\
        \caption{The quantum modified trajectories follow similar paths, as the classic trajectory, to the
        singularity, with $M=\frac{1}{3}$ and $A=0.0009$.}
        \label{d-01}
    \end{center}
\end{figure}

\begin{figure}
    \begin{center}
         \includegraphics[width=0.9\textwidth]{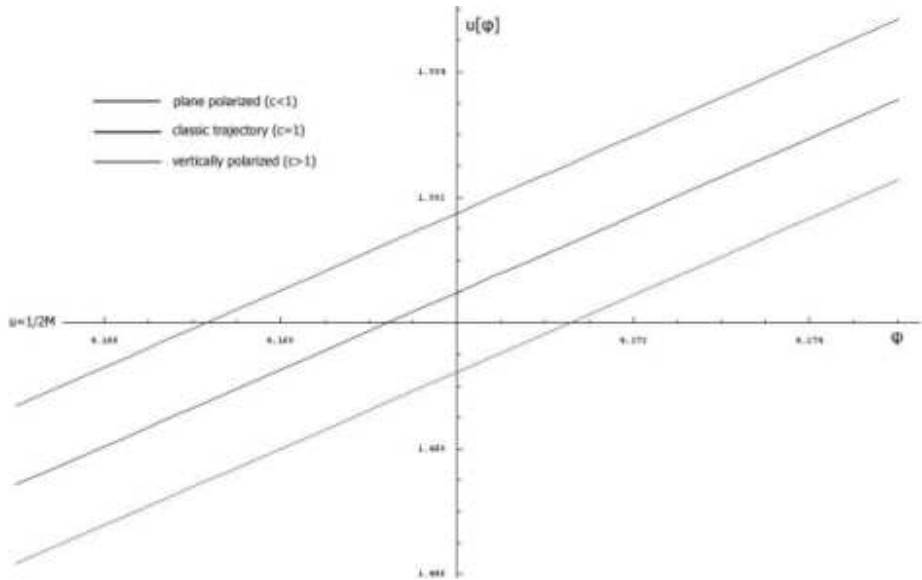}\\
        \caption{The classic and quantum modified trajectories crossing the event horizon $u=2M$, with $M=\frac{1}{3}$ and $A=0.0009$.}
        \label{d-01-zoom}
    \end{center}
\end{figure}

In order to construct quantum modified trajectories, which go past
the critical orbit and fall into the singularity, we require the
impact parameters:
    \begin{equation}
\frac{1}{D^2}=\frac{1}{(\frac{\sqrt{3}}{u_0\sqrt{(1\pm A
Mu_0)}}-\frac{1}{10})^2} \rightarrow
    D=\frac{\sqrt{3}}{u_0}\mp \frac{\sqrt{3}A M}{2}-\frac{1}{10}+\mathcal{O}(A)
    \label{d-less-crit}
    \end{equation}
where, as before, $+$ is for vertical polarization and $-$ is planar
polarization, in $1/D^2$.  For these impact parameters we
numerically solved Eqn.~(\ref{genreal equation of motion dependent
on phi}), and in Fig.~\ref{d-01} we can see that all the
trajectories follow a classic type path into the
singularity\footnote{The quantum modifications to the classic
trajectories are very small, and even if we use the hugely
exaggerated value of $A=0.00009$, as was used in Figs.~\ref{general
critical trajectories large} and \ref{general critical trajectories
small}, the modification is hardly visible. So, in order to magnify
the quantum correction even more we used $A=0.0009$.}.  However,
near the singularity you can see the splitting of the orbits as they
tend to $u\rightarrow\infty$. In Fig.~\ref{d-01-zoom} we have shown
a magnified view of the point where the trajectories cross the event
horizon.  In this figure it can be seen that the planar polarized
photon ($c<1$) crosses the event horizon at a point before the
classic trajectory and the vertically polarized photon ($c>1$)
crosses it at a point after the classic trajectory.  This makes
sense, as the planar polarized photons are pushed towards the black
hole and vertically polarized trajectories are pushed out, the
vertically polarized ones must spiral further around the black hole
to reach the event horizon compared to the classic or the planar
polarized trajectories.
\subsection{Fixed Event Horizon}
\begin{figure}
     \begin{center}
        \includegraphics[width=0.9\textwidth]{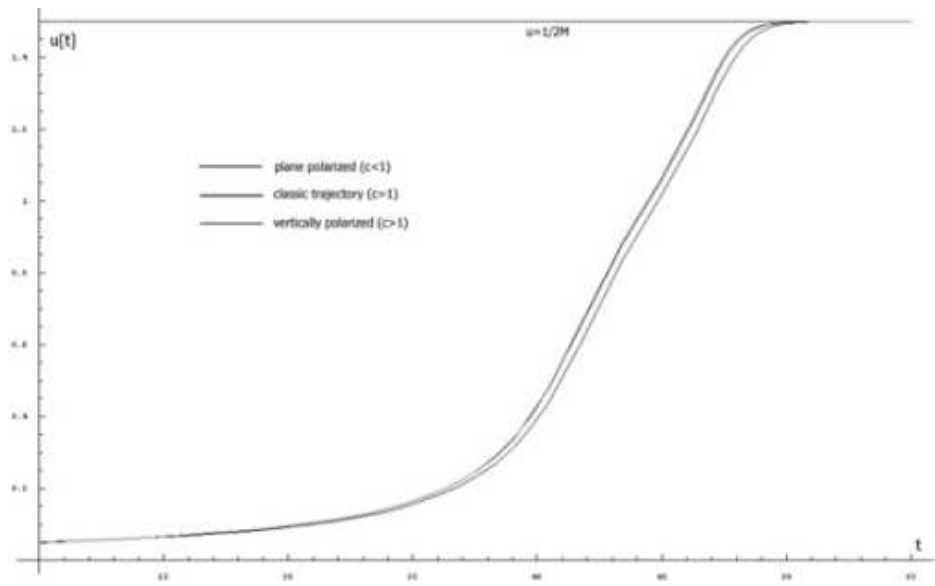}\\
        \caption{The quantum modified trajectories follow similar paths, as the classic trajectory, to the
        event horizon, with $M=\frac{1}{3}$ and $A=0.0009$.}
        \label{u(t)d-01}
    \end{center}
\end{figure}

\begin{figure}
    \begin{center}
         \includegraphics[width=0.9\textwidth]{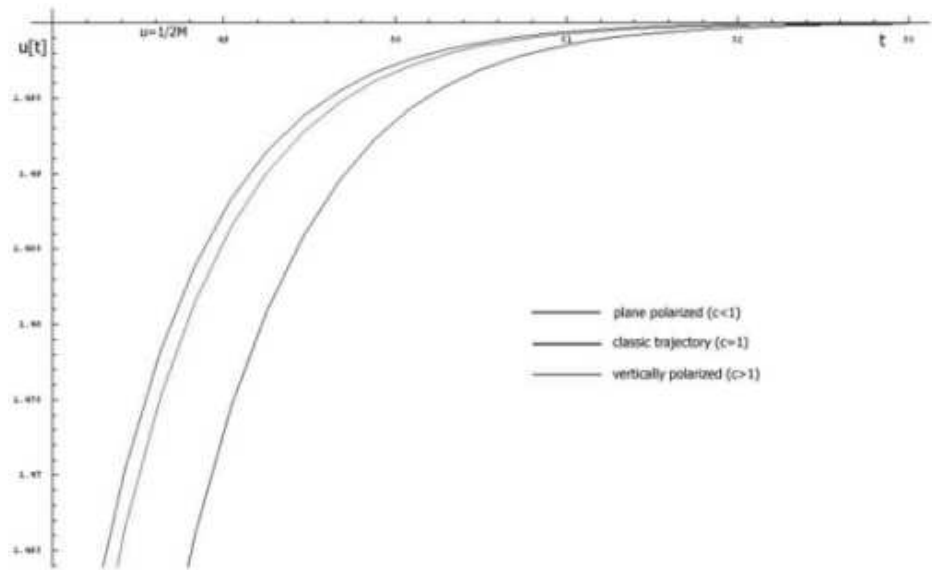}\\
        \caption{The classic and quantum modified trajectories reach the event horizon at differing times, with $M=\frac{1}{3}$ and $A=0.0009$.}
        \label{u(t)d-01-zoom}
    \end{center}
\end{figure}

To demonstrate the fact that the event horizon remains fixed at
$u=1/2M$ we applied the impact parameters given by
(\ref{d-less-crit}) to Eqn.(\ref{genreal equation of motion
dependent on t}).  Again, solving numerically we found that the
trajectories tend to the event horizon in the classic way,
Fig.~\ref{u(t)d-01}, however, they are again slightly shifted.  In
Fig.~\ref{u(t)d-01-zoom}, a close up of the point where the
trajectories tend to the event horizon at $u=1/2M$, you can clearly
see that the quantum modified orbits tend to the horizon before the
classic orbit.  This can be understood by the fact that the planar
polarized trajectory is pushed towards the black hole, hence it has
less of a distance to propagate before it reaches the horizon, and
even though the vertically polarized trajectory is pushed outwards,
the fact that it has a faster velocity than $c=1$ it reaches the
event horizon before the classic trajectory.
The other more important thing we can note from this calculation is
that the event horizon remains fixed at the classic point $u=1/2M$,
due to the fact that, as was stated earlier, the quantum correction
has no effect on photon momentum normal to the horizon.  This means
even though the quantum correction implies velocities greater than
the speed of light, no photons can escape from the event horizon, so
the black hole remains black.

%% file: part1-chapter5.tex
\chapter{Quantum Modified Schwarzschild Metric}
\label{Quantum Modified Schwarzschild Metric} Now, using our
previous results, we can construct a new metric that encompasses the
quantum corrections due to vacuum polarization. This metric should
then be a sum of the classic Schwarzschild metric and the
polarization dependent quantum corrections we derived in
Eqns.~(\ref{correction vertical}) and (\ref{correction planar}):
\begin{eqnarray}
\mathcal{G}_{\mu\nu}k^{\mu}k^{\nu}&=&(g_{\mu\nu}+\mathfrak{g}_{\mu\nu})k^{\mu}k^{\nu}=0\nonumber\\
&=&(g_{\mu\nu}-\frac{8c_{\alpha}}{m_e^2}
R_{\mu\eta\nu\lambda}a^{\eta}a^{\lambda})k^{\mu}k^{\nu}=0
\end{eqnarray}
\section{Construction of the Metric}
Using the results in Sec.~\ref{quantum mod section} we can write the
first components of the quantum modified metric $\mathcal{G}_{tt}^1$
and $\mathcal{G}_{tt}^2$, for vertical and planar polarization
respectively, as:
\begin{eqnarray}
\mathcal{G}_{tt}^1&=&g_{tt}-\frac{8c_{\alpha}}{m_e^2}R_{t b t
d}a_1^b
a_1^d \nonumber\\
&=&F-\frac{8c_{\alpha}}{m_e^2}(-\frac{M
F}{r})(\frac{1}{r^2})\nonumber\\
&=&F+\frac{A M D^2}{9}F u^3\\
\mathcal{G}_{tt}^2&=&g_{tt}-\frac{8c_{\alpha}}{m_e^2}R_{t b t
d}a_2^b
a_2^d \nonumber\\
&=&F-\frac{8c_{\alpha}}{m_e^2}[\frac{2M}{r^3}(\frac{DF}{r})^2 - (\frac{MF}{r})(\frac{\sqrt{1-\frac{D^2F}{r^2}}}{r})^2]\nonumber\\
&=&F-\frac{A M D^2}{9}(3 D^2 F^2 u^5-Fu^3)
\end{eqnarray}
where, in the last lines of each component, we have used
Eqn~(\ref{constant A}) to replace the constants with $A$, and we've
made the transformation $r\rightarrow1/u$. Now, doing this for all
the other components we can construct the metrics for vertical and
planar polarizations:
\begin{itemize}
    \item Vertical Polarization: $a_1^{\mu}=u(0,0,1,0)$
        \begin{equation}
\mathcal{G}^1_{\mu\nu}=\left(
                         \begin{array}{cccc}
                           F-B F u^3 & 0 & 0 & 0 \\
                           0 & -\frac{1}{F}+\frac{B u^3}{ F} & 0 & 0 \\
                           0 & 0 & -\frac{1}{u^2} & 0 \\
                           0 & 0 & 0 & -\frac{1}{u^2}-B u\\
                         \end{array}
                       \right)
                       \label{quantum metric vertical}
\end{equation}
    \item Planar Polarization: $a_2^{\mu}=u(0,D F,0,D \frac{du}{d\phi})$
    \begin{equation}
\mathcal{G}^2_{\mu\nu}=\left(
                         \begin{array}{cccc}
                           F+B F u^3(1 -3D^2 F u^2) & 0 & 0 & 0 \\
                           0 & -\frac{1}{F}-\frac{B u^3 (1-D^2 F u^2)}{F} & 0 & Bu^3 D \frac{d u}{d\phi} \\
                           0 & 0 & -\frac{1}{u^{2}} & 0 \\
                           0 & Bu^3 D \frac{d u}{d\phi} & 0 & -\frac{1}{u^2}-B D^2 F u^3\\
                         \end{array}
                       \right)
                         \label{quantum metric planar}
\end{equation}
\end{itemize}
where $B=\frac{AM D^2}{9}$.  These are now the relevant quantum
modified Schwarzschild metrics for vertical and planar
polarizations\footnote{ In both the metric, ${G}^2_{\mu\nu}$, and
vector, $a_2^{\mu}$, we have made the substitution $r\rightarrow
1/u$ and $ \frac{du}{d\phi}=-\frac{1}{D}\sqrt{1- D^2 F u^2}$}.
\section{Dynamics with the Quantum Modified Metric} We can now
derive the same equations of motion as (\ref{genreal equation of
motion dependent on phi}), but now we can do it simply with the
quantum modified metrics.  Using the general wave vector in a plane:
\begin{eqnarray}
k^{\mu}&=&(\dot{t},\dot{r},\dot{\theta},\dot{\phi})\nonumber\\
&=&(\frac{J}{D}F^{-1},-J\frac{du}{d\phi},0,J u^2)
\end{eqnarray}
where we have transformed to $r\rightarrow 1/u$, and we substitute
for $\dot{t}$ and $\dot{\phi}$ as before.  Applying this wave vector
to (\ref{quantum metric vertical}) and(\ref{quantum metric planar})
we find:
\begin{eqnarray}
0&=&\mathcal{G}^1_{\mu\nu}k^{\mu}k^{\nu}\nonumber\\
 &\Rightarrow& (\frac{du}{d\phi})^2=\frac{1}{D^2}-Fu^2+
 (\frac{D^2A}{3})Fu^5
 \end{eqnarray}
 for vertical polarization, and
 \begin{eqnarray}
 0&=&\mathcal{G}^2_{\mu\nu}k^{\mu}k^{\nu}\nonumber\\
 &\Rightarrow& (\frac{du}{d\phi})^2=\frac{1}{D^2}-Fu^2- (\frac{D^2A}{3})Fu^5
\end{eqnarray}
for planar polarization.  These are identical to the equations we
derived in Sec.~\ref{quantum mod section}.
\subsubsection{Radial Geodesics}
We can now show that the metric for vertical polarization is
consistent with the fact that radially projected photon trajectories
are not modified. By using the vertical polarization metric,
(\ref{quantum metric vertical}), and a general radial wave vector,
($k^{\mu}=(\dot{t},\dot{r},0,0)$, we find:
\begin{eqnarray}
0&=&\mathcal{G}^1_{\mu\nu}k^{\mu}k^{\nu}\nonumber\\
\Rightarrow \qquad \frac{dt}{dr}&=&\pm\frac{1}{F}
\end{eqnarray}
which is identical to the classic radial geodesic equation,
(\ref{rearranged dr/dt J=0}).  However, the same is not true for the
planar polarization metric.  Due to our derivation of the
polarization vectors in Sec.~\ref{polarization vectors}, we
constructed a very general vertical polarization vector and then
normalized it; however, the one for planar polarization was
constructed for the case where $\frac{d\phi}{d\tau}\neq0$, as can be
seen in Eqns.~(\ref{dr/dtau}) and (\ref{wave and polarization});
therefore the planar polarization vector has $\phi$ dependence
"mixed" into it through the substitution of $\dot{\phi}$:
\begin{eqnarray}
(\frac{dr}{d\tau})^2&=&F^2(\frac{dt}{d\tau})^2-r^2F(\frac{d\phi}{d\tau})^2\nonumber\\
&=&(E^2-\frac{J^2 F}{r^2})^{\frac{1}{2}}=E(1-\frac{D^2
F}{r^2})^{\frac{1}{2}}\nonumber\\
\end{eqnarray}
We can note that it's a result of the substitution
$\frac{d\phi}{d\tau}=\frac{J}{r^2}$, in the polarization vectors,
that we acquire an extra parameter $D$ in our quantum correction
(apart from the one introduced through normalization, which was
incorporated into $B=A D^2/9$). We can then say that if we need to
use the metric for a radial trajectory with $\frac{d\phi}{d\tau}=0$
we can just set $D=0$, as this would lead to the removal of the
$\phi$ dependence. In this way, the planar polarized quantum
modified metric for radial trajectories is:
\begin{equation}
\mathcal{G}(\textsl{$\phi$=0})^2_{\mu\nu}=\left(
                         \begin{array}{cccc}
                           F+B F u^3 & 0 & 0 & 0 \\
                           0 & -\frac{1}{F}-\frac{B u^3 }{F} & 0 & 0 \\
                           0 & 0 & -\frac{1}{u^{2}} & 0 \\
                           0 & 0& 0 & 0\\
                         \end{array}
                       \right)
\end{equation}
Then applying this to a general radial wavevector, we again find the
classic radial geodesic equation:
\begin{eqnarray}
0&=&\mathcal{G}(\textsl{$\phi$=0})^2_{\mu\nu}k^{\mu}k^{\nu}\nonumber\\
\Rightarrow \qquad \frac{dt}{dr}&=&\pm\frac{1}{F}
\end{eqnarray}
Therefore, we have metrics for the Schwarzschild space time that
incorporate quantum corrections due to vacuum polarization; and
these metrics are consistent with classic results.

%% file: part1-chapter6.tex
\chapter{Summary}
\label{part1-summary} For Schwarzschild Spacetime we showed, in the
orthonormal frame, that, due to quantum corrections, the stable
circular orbit at $u_0=1/3M$ splits depending on the polarization of
the photon, and the new modified circular orbits are given by the
classical orbit plus a correction term to first order in a
dimensionless constant $A$: $u=u_0+ Au_1$, where
$A=\frac{24c_{\alpha}E_k^2}{J^2m_e^2}$ and has an order of
$10^{-32}$. As stated by the polarization sum rule this splitting of
the critical orbit is equal in magnitude but opposite in sign for
the two polarizations. We found that the vertically polarized photon
(c>1) is pushed out by a correction of $u_1=-\frac{M}{6}u_0^2$,
while the planar polarized photon is pulled in by a correction of
$u_1=\frac{M}{6}u_0^2$. It was also found that this orbit shift also
requires an appropriate modification of the classic impact parameter
$\frac{1}{D_0^2}=\frac{u_0^2}{3}$. This modification, for the
quantum corrected orbits, took the form
$\frac{1}{D^2}=\frac{1}{D_0^2}+A D_1$, and again as for the orbit
shift, the change was equal in magnitude but opposite in sign for
the two polarizations: $D_1=\frac{Mu_0^3}{3}$ for planar
polarization and $D_1=-\frac{Mu_0^3}{3}$ for vertical polarization.
Using this information, for the splitting of circular orbits, we
then constructed the quantum corrected general equations of motion
for the Schwarzschild spacetime.  Using these equation of motion it
was shown that a photon starting at $u=0$, with the appropriate
critical impact parameter, tends to the critical orbit associated
with that impact parameter - and the trajectory follows a similar
path to the classic case.

We then went on to show that, using the general quantum corrected
equations of motion, a photon projected towards the black hole with
an impact parameter less than the critical value falls into the the
singularity, in terms of the angular distance ($\phi$). Although the
photons follow a classic type path into the singularity, the
trajectories are slightly shifted according to polarization. The
planar polarized photon crosses the horizon before the classic
trajectory, and the vertically polarized one crosses it after, this
corresponds to the fact that planar polarized photons are pulled
towards the black hole and vertically polarized ones are pushed
away, hence the vertically polarized ones need to go a further
angular distance to reach the event horizon.

In terms of coordinate time ($t$) we found that the photon
trajectories tend to the event horizon. Therefore, the quantum
corrections do not shift the classic event horizon from
$u=\frac{1}{2M}$, which corresponds to the horizon theorem.  Also,
we found that, although the quantum corrected orbits follow a
classic type path to the event horizon, the point at which they hit
the horizon is, again, slightly shifted depending on polarization.
However, this time both polarizations hit the horizon before the
classic trajectory. The planar polarized photon tends to arrive at
the horizon first, then the vertically polarized one, and finally
the classic photon.  This could correspond to the fact that the
planar polarized photon, although it has a velocity lower than the
speed of light, has less of a distance to go, as its trajectory is
pulled towards the black hole.  For the vertically polarized case,
even though it has a faster than light velocity, it has a further
distance to go to reach the horizon as its trajectory is pushed away
from the black hole.

Having determined the equations of motion, with the quantum
correction, we then went on to construct a Schwarzschild metric that
incorporates the quantum correction:
$\mathcal{G}_{\mu\nu}=(g_{\mu\nu}+\mathfrak{g}_{\mu\nu})$, where the
correction: $\mathfrak{g}_{\mu\nu}$ was again first order in $A$. We
showed that with this metric and a general photon wave vector,
$k^{\mu}$, we obtain the quantum modified equations of motion, as
before.  Also, this new metric confirms the horizon theorem, that
is, when we use a wave vector indicating a radially projected photon
we obtain the classic equation of motions.  This was fine for
vertical polarization, however, in the planar polarization case we
had a problem; we had previously used a substitution that mixed a
"hidden" radial angle, $\phi$, into our polarization vector (as in
general orbits the planar polarization depends on $\phi$). By
tracing back to the origin of this substitution we found that, in
order to study radial trajectories, we need to set $D=0$, this then
removes the $\phi$ dependency; this then gives us the classic
equation of motion for planar polarized radially projected photons.

So in conclusion, after studying the dynamics of null trajectories
in  Schwarzschild spacetime we derived the polarization dependent
photon trajectories to first order in the constant $A$ (which is
dependent on the fine structure constant, the mass of the star, mass
of the electron, and the energy of the photon). We then incorporated
these modification into a general quantum modified metric, which
could also be used to derive the general quantum modified equations
of motion.  Also, the results of this work coincide with the
conditions of the horizon theorem and the polarization sum rule.

%% file: app1.tex
\appendix{}
\chapter{Schwarzschild Spacetime}
\label{app schwart} The 'Schwarzschild geometry' describes a
spherically symmetric spacetime outside a star, and its properties
are determined by one parameter, the mass M.  The Schwarzschild
metric, in spherical polar coordinates, takes the form:
\begin{equation}
        g_{\mu\nu}=\left(%
        \begin{array}{cccc}
        F & 0 & 0 & 0 \\
        0 & -F^{-1} & 0 & 0 \\
        0 & 0 & -r^2 & 0 \\
        0 & 0 & 0 & -r^2\sin(\theta) \\
        \end{array}%
        \right)
    \end{equation}
This type of spacetime geometry is said to be static, due to the
fact that (i) all metric components are independent of $t$, and (ii)
the geometry is unchanged by time reversal, $t\rightarrow
-t$\footnote{A space time with property (i) but not necessarily (ii)
is said to be stationary i.e. a rotating star/black hole}.
\section{Geodesic Equations}
The equations of motion in Schwarzschild spacetime can be derived by
using the covariant form of Newton's second law of motion,
\begin{eqnarray}
F^{\mu}&=&\frac{D p^{\mu} }{D\tau}\nonumber\\
&=&\frac{dp^{\mu}}{d\tau}+\Gamma^{\mu}_{\nu\rho}\frac{dx^{\nu}}{d\tau}p^{\rho}
\end{eqnarray}
where we have $p^{\mu}=m\frac{dx^{\mu}}{d\tau}$, and for free fall
we have $F^{\mu}=0$, then the equations of motion are:
\begin{equation}
0=\frac{d^2x^{\mu}}{d\tau^2}+\Gamma^{\mu}_{\nu\rho}\frac{dx^{\nu}}{d\tau}\frac{dx^{\rho}}{d\tau}
\end{equation}
Now, using this and the appropriate Christoffels we are able to
determine the equations of motion.  Alternatively, by using the
Schwarzschild Lagrangian
    \begin{equation}
        L=\frac{1}{2}[ (1-\frac{2M}{r}) \dot{dt} ^{2} - (1-\frac{2M}{r}) ^{-1} \dot{dr}^{2}-r^{2} \dot{d\theta}^{2}-(r^{2}
        \sin^{2} \theta) \dot{d\phi}^{2}]
        \label{schwarzschild lagrangian}
    \end{equation}
we find
\begin{eqnarray}
        \frac{\partial L}{\partial t}&=&0 \qquad \frac{\partial L}{\partial \dot{t}}=(1-\frac{2M}{r})\dot{t}\nonumber\\
        \frac{d}{d \lambda}\frac{\partial L}{\partial \dot{t}}&=&(1-\frac{2M}{r})\ddot{t}+\frac{2M}{r^{2}}
        \dot{r}\dot{t}\nonumber\\
\end{eqnarray}
\begin{eqnarray}
        \frac{\partial L}{\partial r}&=&\frac{M}{r^2}\dot{t}^2+\frac{M}{r}^{2}(1-\frac{2M}{r})^{-2}\dot{r}^{2} -r
        \dot{\theta}^{2}-r \dot{\phi}^{2} \sin^{2}\theta\nonumber\\
        \frac{\partial
        L}{\partial\dot{r}}&=&-(1-\frac{2M}{r})^{-1}\dot{r}\nonumber\\
        \frac{d}{d \lambda} \frac{\partial L}{\partial \dot{r}}&=&-(1-\frac{2M}{r})^{-1}\ddot{r}+\frac{2M}{r^{2}}(1-
        \frac{2M}{r})^{-2}\dot{r}\dot{r}\nonumber\\
\end{eqnarray}
\begin{eqnarray}
        \frac{\partial L}{\partial \theta}&=&-r^{2} \sin\theta \cos\theta \dot\phi^{2} \qquad
        \frac{\partial L}{\partial
        \dot{\theta}}=-r^{2}\dot{\theta}\nonumber\\
        \frac{d}{d \lambda}\frac{\partial L}{\partial
        \dot{\theta}}&=&-2r\dot{r}\dot{\theta}-r^{2}\ddot{\theta}\nonumber\\
\end{eqnarray}
\begin{eqnarray}
        \frac{\partial L}{\partial \phi}&=&0 \qquad
        \frac{\partial L}{\partial \dot{\phi}}=-r^{2}\sin^{2}\theta
        \dot{\phi}^{2}\nonumber\\
        \frac{d}{d\lambda}\frac{\partial L}{\partial \dot{\phi}}&=&-2r\dot{r}\sin^2\theta \dot{\phi}-2r^2\sin\theta \cos\theta
                 \dot{\phi}-r^2\sin^2\theta \ddot{\phi}\nonumber\\
        \end{eqnarray}
and substituting into the Euler equation
    \begin{equation}
        \frac{\partial L}{\partial x^{\mu}}-\frac{d}{d\lambda}\frac{\partial L}{\partial \dot{x}^{\mu}}=0
        \label{euler equation}
    \end{equation}
 we have the equations of motion for Schwarzschild spacetime
    \begin{eqnarray}
  \label{equation of motion t}
        0&=&(1-\frac{2M}{r})\ddot{t}+\frac{2M}{r^2}\dot{r}\dot{t}=0\\
  \label{equation of motion r}
        0&=&\frac{M}{r^2}\dot{t}^2-r\dot{\theta}^{2}-r \dot{\phi}^{2}
        \sin^{2}\theta+(1-\frac{2M}{r})^{-1}\ddot{r}\nonumber\\
        &-&\frac{M}{r^{2}}(1-\frac{2M}{r})^{-2}\dot{r}^{2}\\
\label{equation of motion theta}
       0&=& -r^{2} \sin\theta \cos\theta
       \dot\phi^{2}+2r\dot{r}\dot{\theta}+r^{2}\ddot{\theta}\\
 \label{equation of motion phi}
        0&=&2r\dot{r}\sin^2\theta \dot{\phi}+2r^2\sin\theta \cos\theta \dot{\phi}+r^2\sin^2\theta
        \ddot{\phi}\\
        \nonumber
    \end{eqnarray}
\section{Riemann Components in the Orthonormal Frame}
\label{orthonormal} The spacetime line interval,
Eqn.~(\ref{schwarzschild line interval}), can be rewritten using
tetrad transformations (discussed in the next section):
    \begin{equation}
        \omega^t=(1-\frac{2M}{r})^{\frac{1}{2}}dt \qquad
        \omega^r=(1-\frac{2M}{r})^{-\frac{1}{2}}dr \qquad
        \omega^\theta=r d\theta \qquad
        \omega^\phi=r \sin(\theta)d\phi
    \label{omega transforms}
    \end{equation}
as
    \begin{equation}
         ds^{2}=(\omega^{t})^{2}-(\omega^{r})^{2}-(\omega^{\theta})^{2}-(\omega^{\phi})^{2}
        \label{transformed omega interval}
    \end{equation}
This now has the Minkowski metric
    \begin{equation}
    g_{\mu\nu}=\left(%
\begin{array}{cccc}
  1 & 0 & 0 & 0 \\
  0 & -1 & 0 & 0 \\
  0 & 0 & -1 & 0 \\
  0 & 0 & 0 & -1 \\
\end{array}%
\right)
        \label{minkoski metric appedix}
    \end{equation}
Noting that the metric can be written as:
    \begin{equation}
        dg_{\mu\nu}=\omega_{\mu\nu}+\omega_{\nu\mu}
    \end{equation}
and using the fact that:
    \begin{equation}
        dg_{\mu\nu}=\frac{\partial g_{\mu\nu}}{\partial
        x^{\alpha}}dx^\alpha=0
    \end{equation}
we then have\cite{chandrasekhar}:
    \begin{equation}
        \omega_{\mu\nu}=-\omega_{\nu\mu}
    \end{equation}
which implies $\omega$ is antisymmetric, i.e. it has only six unique
components and $\omega_{\mu\nu}=0$ for $\mu=\nu$; and we have the
conditions:
    \begin{equation}
        \omega^0_i=\omega^i_0 \qquad \omega^i_j=-\omega^j_i
    \end{equation}
We can now write the exterior derivatives\footnote{The exterior
derivative: $d=dx^{\mu}\frac{\partial}{\partial x^{\mu}}$ acting on
a 1-form $A^{\nu}=A(x)dx^{\nu}$ gives $dA^{\nu}=dx^{\mu}\wedge
dx^{\nu}\frac{\partial A(x)}{\partial x^{\mu}}$, where
$dx^{\mu}\wedge dx^{\nu}=0$ for $\mu=\nu$ } of $\omega^{\mu}$
    \begin{eqnarray}
  \label{exterior t}
        d\omega^t&=&\frac{1}{2}(1-\frac{2M}{r})^{- \frac{1}{2}}(\frac{2M}{r^2})dr\wedge
        dt\nonumber\\
        &=&\frac{M}{r^2}(1-\frac{2M}{r})^{-\frac{1}{2}}\omega^r\wedge\omega^t\\
     \label{exterior r}
        d\omega^r&=&0\\
  \label{exterior theta}
        d\omega^{\theta}&=&dr\wedge d\theta=\frac{1}{r}(1-\frac{2M}{r})^{\frac{1}{2}}\omega^r \wedge
        \omega^{\theta}\\
 \label{exterior phi}
        d\omega^{\phi}&=&\sin(\theta) dr\wedge
        d\phi+r\cos(\phi)d\theta \wedge
        d\phi\nonumber\\
        &=&\frac{1}{r}(1-\frac{2M}{r})^{\frac{1}{2}}\omega^r \wedge
        \omega^{\phi}+\frac{\cot(\theta)}{r} \omega^\theta \wedge
        \omega^{\phi}\\
        \nonumber
    \end{eqnarray}
Now, using Cartan's equation
        \begin{equation}
            d\omega^\mu=\omega^\alpha \wedge
            \omega^\mu_\alpha+\Omega^\mu
        \end{equation}
    with zero torsion $(\Omega^\mu=0)$, we can write
\begin{eqnarray}
   \label{domega t}
        d\omega^t&=&\omega^r\wedge \omega^t_r+\omega^\theta \wedge
        \omega^t_\theta+\omega^\phi \wedge \omega^t_\phi\\
      \label{domega_r}
        d\omega^r&=&\omega^t\wedge
        \omega^r_t+\omega^\theta\wedge\omega^r_\theta+\omega^\phi\wedge\omega^r_\phi\\
   \label{domega theta}
        d\omega^\theta&=&\omega^t\wedge\omega^\theta_t+\omega^r\wedge\omega^\theta_r+\omega^\phi\wedge\omega^\theta_\phi\\
     \label{domega phi}
        d\omega^\phi&=&\omega^t\wedge\omega^\phi_t+\omega^r\wedge\omega^\phi_r+\omega^\theta\wedge\omega^\phi_\theta\\
\nonumber
    \end{eqnarray}
Comparing Eqns.~(\ref{exterior t})-(\ref{exterior phi})
    with Eqns.~(\ref{domega t})-(\ref{domega phi}) we find the
    six unique components of $\omega^\mu_\nu $ as
    \begin{eqnarray}
        \omega^t_r&=&\frac{M}{r^2}(1-\frac{2M}{r})^{-\frac{1}{2}}\omega^t=\frac{M}{r^2}dt=\omega^r_t\\
        \omega^\theta_r&=&\frac{1}{r}(1-\frac{2M}{r})^{\frac{1}{2}}\omega^\theta=(1-\frac{2M}{r})^{\frac{1}{2}}d\theta=-\omega^r_\theta\\
        \omega^\phi_r&=&\frac{1}{r}(1-\frac{2M}{r})^{\frac{1}{2}}\omega^\phi=\sin(\theta)(1-\frac{2M}{r})^{\frac{1}{2}}d\phi=-\omega^r_\phi\\
        \omega^\phi_\theta&=&\frac{\cot(\theta)}{r}\omega^\phi=\cos(\theta)d\phi=-\omega^\theta_\phi\\
        \omega^\theta_t&=&\omega^t_\theta=0\\
        \omega^\phi_t&=&\omega^t_\phi=0\\
        \nonumber
    \end{eqnarray}
Now, using:
    \begin{equation}
        R^\mu_\nu=d\omega^\mu_\nu+\omega^\mu_\alpha\wedge\omega^\alpha_\nu
    \end{equation}
we can write the six independent components of the Riemann tensor:
\begin{eqnarray}
        R^t_r&=&d\omega^t_r+\omega^t_\alpha\wedge\omega^\alpha_r=d\omega^t_r+\omega^t_\theta\wedge\omega^\theta_r+\omega^t_\phi\wedge\omega^\phi_r\nonumber\\
        &=&-\frac{2M}{r^3}dr\wedge
        dt=-\frac{2M}{r^3}\omega^r\wedge\omega^t\nonumber\\
         &\Rightarrow& R^{t}_{rrt}=R_{trrt}=-\frac{2M}{r^3}\nonumber\\
 \label{R t r r t}
\end{eqnarray}
\begin{eqnarray}
         R^{\theta}_r&=&d \omega^{\theta}_r+ \omega^{\theta}_\alpha \wedge \omega^{\alpha}_r
         =d \omega^{\theta}_r+\omega^{\theta}_t \wedge \omega^t_r
         +\omega^{\theta}_\phi\wedge\omega^{\phi}_r\nonumber\\
         &=&\frac{1}{2}(1-\frac{2M}{r})^{-\frac{1}{2}}\frac{2M}{r^2}dr \wedge d \theta=
         \frac{M}{r^3} \omega^r \wedge \omega^\theta\nonumber\\
        &\Rightarrow& R^{\theta}_{rr\theta}=-R_{\theta
        rr\theta}=\frac{M}{r^3}\nonumber\\
\label{R theta r r theta}
        \end{eqnarray}
\begin{eqnarray}
       R^{\phi}_r&=&d \omega^{\phi}_r+ \omega^{\phi}_\alpha \wedge \omega^{\alpha}_r=d \omega^{\phi}_r+\omega^{\phi}_t \wedge \omega^t_r
         +\omega^{\phi}_\theta\wedge\omega^{\theta}_r\nonumber\\
         &=&\cos(\theta)(1-\frac{2M}{r})^{\frac{1}{2}}d\theta\wedge
         d\phi+\sin(\theta)\frac{M}{r^2}(1-\frac{2M}{r})^{-\frac{1}{2}}dr\wedge
         d\phi\nonumber\\
         &+&\cos(\theta)(1-\frac{2M}{r})^{1}{2}d\phi\wedge
         d\theta\nonumber\\
         &=&\frac{\cot(\theta)}{r^2}(1-\frac{2M}{r})^{\frac{1}{2}}
         \omega^\theta\wedge \omega^\phi +
         \frac{M}{r^3}\omega^r \wedge
         \omega^\phi\nonumber\\
         &+&\frac{\cot(\theta)}{r^2}(1-\frac{2M}{r})^{\frac{1}{2}}\omega^\phi \wedge
         \omega^\theta=\frac{M}{r^3}\omega^r\wedge \omega^\phi\nonumber\\
        &\Rightarrow& R^{\phi}_{rr\phi}=-R_{\phi
        rr\phi}=\frac{M}{r^3} \nonumber\\
   \label{R phi r r phi}
        \end{eqnarray}
\begin{eqnarray}
         R^{\phi}_\theta&=&d \omega^{\phi}_\theta+ \omega^{\phi}_\alpha \wedge \omega^{\alpha}_\theta=d \omega^{\phi}_\theta+\omega^{\phi}_t \wedge
         \omega^t_\theta+\omega^{\phi}_r\wedge\omega^r_\theta\nonumber\\
         &=&-\sin(\theta)d\theta\wedge
         d\phi-\sin(\theta)(1-\frac{2M}{r})d\phi\wedge
         d\theta\nonumber\\
         &=&-\frac{1}{r^2}\omega^\theta\wedge \omega^\phi +
         \frac{1}{r^2}(1-\frac{2M}{r})\omega^\theta \wedge
         \omega^\phi=-\frac{2M}{r^3}\omega^\theta\wedge\omega^\phi\nonumber\\
         &\Rightarrow&
         R^\phi_{\theta\theta\phi}=-R_{\phi\theta\theta\phi}=-\frac{2M}{r^3} \nonumber\\
 \label{R phi theta theta phi}
        \end{eqnarray}
\begin{eqnarray}
         R^{\theta}_t&=&d \omega^{\theta}_t+ \omega^{\theta}_\alpha \wedge \omega^{\alpha}_t=
         d \omega^{\theta}_t+\omega^{\theta}_r \wedge \omega^r_t
         +\omega^{\theta}_\phi\wedge\omega^{\phi}_t\nonumber\\
         &=&\frac{M}{r^2}(1-\frac{2M}{r})^{\frac{1}{2}}d\theta \wedge dt
         =\frac{M}{r^3} \omega^\theta \wedge \omega^t\nonumber\\
         & \Rightarrow &R^{\theta}_{t\theta t}=-R_{\theta t\theta
         t}=\frac{M}{r^3} \nonumber\\
\label{R theta t theta t}
         \end{eqnarray}
\begin{eqnarray}
         R^{\phi}_t&=&d \omega^{\phi}_t+ \omega^{\phi}_\alpha \wedge \omega^{\alpha}_t=
         d \omega^{\phi}_t+\omega^{\phi}_r \wedge \omega^r_t
         +\omega^{\phi}_\theta\wedge\omega^{\theta}_t\nonumber\\
         &=&\frac{M}{r^2}(1-\frac{2M}{r})^{\frac{1}{2}}\sin(\theta)d\theta \wedge dt
         =\frac{M}{r^3} \omega^\phi \wedge \omega^t\nonumber\\
        &\Rightarrow& R^{\phi}_{t\phi t}=-R_{\phi t\phi
        t}=\frac{M}{r^3}\nonumber\\
  \label{R phi t phi t}
        \end{eqnarray}
\section{Tetrad Transformation}
\label{tetrad} In Eqn.~(\ref{omega transforms}) we transformed from
the coordinate frame to an orthonormal frame. This is achieved by
using the tetrad of the form:
       \begin{equation}
            e^{a}_{\mu}=\left(%
            \begin{array}{cccc}
            F^{\frac{1}{2}} & 0 & 0 & 0 \\
            0 & F^{-\frac{1}{2}} & 0 & 0 \\
            0 & 0 & r & 0 \\
            0 & 0 & 0 & r\sin(\theta) \\
            \end{array}%
            \right)
            \label{tetrad}
    \end{equation}
This takes a 4-vector, given in the coordinate frame, and maps it to
the equivalent in the orthonormal frame.  We can then determine the
inverse tetrad, i.e. a tetrad that takes vectors in the orthonormal
frame and maps them to the coordinate frame.  We find this by using
the inverse of (\ref{tetrad}),
    \begin{equation}
        (e^{-1})^{\mu}_{a}=\left(%
        \begin{array}{cccc}
        F^{-\frac{1}{2}} & 0 & 0 & 0 \\
        0 & F^{\frac{1}{2}} & 0 & 0 \\
        0 & 0 & \frac{1}{r} & 0 \\
        0 & 0 & 0 & \frac{1}{r\sin(\theta)} \\
        \end{array}%
        \right),
        \label{inverse tetrad}
    \end{equation}
which has the property:
    \begin{equation}
        e^{a}_{\mu}(e^{-1})^{\mu}_{b}=\delta^a_b
    \end{equation}
Therefore, (\ref{inverse tetrad}) takes vectors in the orthonormal
frame and maps them to the coordinate frame.
\section{Riemann Components in the Coordinate Frame}
In Sec.~\ref{orthonormal} the unique components of the Riemann
tensor were calculated in the orthonormal frame.  In order to
determine the components in the coordinate basis we will use the
transformation tetrad given in Eqn.~\ref{inverse tetrad}.  Before we
can us the tetrad, (\ref{inverse tetrad}), we need to adjust it so
it is able to take covectors in the orthonormal frame, rather than
vectors:
    \begin{equation}
        (e^{-1})^{\mu}_ag_{\mu\nu}\eta^{ab}=(e^{-1})_\nu^{b}=\left(%
        \begin{array}{cccc}
          F^{\frac{1}{2}} & 0 & 0 & 0 \\
          0 & F^{-\frac{1}{2}} & 0 & 0 \\
          0 & 0 & r & 0 \\
          0 & 0 & 0 & r \\
          \end{array}%
            \right),
            \label{modified inverse tetrad}
    \end{equation}
which is just Eqn.~(\ref{tetrad}).  Then using this we have:
    \begin{displaymath}
        \Rightarrow \qquad A_\mu(\textrm{coordinate
        frame})=(e^{-1})_{\mu}^aA_a(\textrm{orthonormal frame})
    \end{displaymath}
Now using this tetrad we can transform the orthonormal Riemann
tensor components as:
        \begin{displaymath}
            R_{\mu\nu\alpha\beta}=(e^{-1})_{\mu}^{a}
            (e^{-1})_{\nu}^{b}(e^{-1})_{\alpha}^{c}(e^{-1})_{\beta}^{d}R_{abcd}
        \end{displaymath}
Now using the components given in Sec.~\ref{orthonormal} we have:
    \begin{eqnarray}
        R'_{trrt}&=&(e^{-1})_{t}^{t}
            (e^{-1})_{r}^{r}(e^{-1})_{r}^{r}(e^{-1})_{t}^{t}R_{trrt}=F^{\frac{1}{2}}
            F^{-\frac{1}{2}}F^{-\frac{1}{2}}F^{\frac{1}{2}}(-\frac{2M}{r^3})=-\frac{2M}{r^3}\nonumber\\
        R'_{\theta r r \theta}&=&(e^{-1})_{\theta}^{\theta}
            (e^{-1})_{r}^{r}(e^{-1})_{r}^{r}(e^{-1})_{\theta}^{\theta}R_{\theta r r \theta}=r F^{-\frac{1}{2}}
            F^{-\frac{1}{2}}r(-\frac{M}{r^3})=-\frac{MF^{-1}}{r}\nonumber\\
        R'_{\phi rr \phi}&=&(e^{-1})_{\phi}^{\phi}
            (e^{-1})_{r}^{r}(e^{-1})_{r}^{r}(e^{-1})_{\phi}^{\phi}R_{\phi r r
            \phi}=r
            F^{-\frac{1}{2}}F^{-\frac{1}{2}}r(-\frac{M}{r^3})=-\frac{MF^{-1}}{r}\nonumber\\
        R'_{\phi \theta \theta \phi}&=&(e^{-1})_{\phi}^{\phi}
            (e^{-1})_{\theta}^{\theta}(e^{-1})_{\theta}^{\theta}(e^{-1})_{\phi}^{\phi}R_{\phi \theta \theta
            \phi}=r r r r(\frac{2M}{r^3})=2Mr\nonumber\\
        R'_{\theta t \theta t}&=&(e^{-1})_{\theta}^{\theta}
            (e^{-1})_{t}^{t} (e^{-1})_{\theta}^{\theta} (e^{-1})_{t}^{t}  R_{\theta t \theta t}=r F^{\frac{1}{2}}
             r F^{\frac{1}{2}}(-\frac{M}{r^3})=-\frac{M
             F}{r}\nonumber\\
        R'_{\phi t \phi t}&=&(e^{-1})_{\phi}^{\phi}
            (e^{-1})_{t}^{t}(e^{-1})_{\phi}^{\phi} (e^{-1})_{t}^{t} R_{\phi
            t \phi} t=r  F^{\frac{1}{2}} r F^{\frac{1}{2}}(-\frac{M}{r^3})=-\frac{M
            F}{r}\nonumber\\
    \end{eqnarray}
Therefore, the Riemann tensor components in the coordinate frame are
given as:
    \begin{eqnarray}
   \label{coordinate frame Riemann components}
        R'_{trrt}&=&-\frac{2M}{r^3} \qquad  R'_{\theta r r
        \theta}=-\frac{MF^{-1}}{r} \qquad R'_{\phi rr
        \phi}=-\frac{MF^{-1}}{r}\nonumber\\
         R'_{\phi \theta \theta
        \phi}&=&2Mr \qquad  R'_{\theta t \theta t}=-\frac{M F}{r}
        \qquad R'_{\phi t \phi t}=-\frac{M F}{r}\nonumber\\
    \end{eqnarray}

%% file: space.bbl
\begin{thebibliography}{99}
\bibitem{drummond} I.T. Drummond and S. Hathrell, Phys. Rev. D22 (1980) 343.
\bibitem{chandrasekhar} S. Chandrasekhar, The Mathematical Theory of
Black Holes, Oxford Science Publications.
\bibitem{weinberg} S. Weinberg, Gravitation and Cosmology, Wiley.
\bibitem{kenyon} I. R. Kenyon, General Relativity, Oxford.
\bibitem{graham2} G. M. Shore, Quantum Gravitational Optics, (2003)
gr-qc/0304059.
\bibitem{schneider} P. Schneider J. Ehlers E. E.
Falco, Gravitationl Lenses, Springer-Verlag.
\bibitem{graham} G. M. Shore, "Faster Than Light" Photons in
Gravitational Fields - Causality, Anomalies and Horizons, (1995)
gr-qc/9504041.
\bibitem{daniels} R. D. Daniels G. M. Shore, Faster Than Light
Photons and Rotating Black Holes, (1995) gr-qc/9508048.
\bibitem{hawking} S. W. Hawking G. F. R. Ellis, The Large Scale
Structure of Spacetime, (1973) Cambridge University Press.
\bibitem{graham3} G. M. Shore, Causality and Superluminal Light,
(2003) gr-qc/0302116.
\end{thebibliography}
